\documentclass[10pt-default]{article}
\usepackage{amsmath}
\usepackage{amssymb}
\usepackage{amsfonts}
\usepackage[dvips]{graphicx}

\input{tcilatex}

\begin{document}

{\huge \ }

{\huge Veneziano Amplitudes, }

{\huge Spin Chains and String Models}

$\ \ \ \ \ \ \ \ \ \ \ \ \ \ \ \ \ \ \ \ \ \ \ \ \ \ \ \ \ \ \ \ \ $

Arkady \ L. Kholodenko\footnote{%
E-mail address: string@clemson.edu}

\textit{375 H.L.Hunter Laboratories, Clemson University,Clemson,}

\textit{SC} 29634-0973, U.S.A.

In a series of recently published papers we reanalyzed the existing
treatments of Veneziano and Veneziano-like amplitudes and the models
associated with these amplitudes. In this work we demonstrate that the
already obtained \ new partition function for these amplitudes can be
exactly mapped into that for the Polychronakos-Frahm (P-F) spin chain model.
This observation allows us to recover many of the existing string-theoretic
models, including the most recent ones.


\ 

\ 

\pagebreak

\ \ \ \ \ \ 

\ \ \ \ \ \ \ \ \ \ \ 

\section{Introduction}

Since time when quantum mechanics (QM) was born (in 1925-1926) two seemingly
opposite approaches for description of atomic and subatomic physics were
proposed respectively by Heisenberg and Schr\"{o}dinger. Heisenberg's
approach is aimed at providing the affirmative answer to the following
question: Is combinatorics of spectra (of observables) \ provides sufficient
\ information about microscopic system so that dynamics of such a system can
be described in terms of known macroscopic concepts? Schrodinger's approach
is exactly opposite and is aimed at providing the affirmative answer to the
following question: Using some plausible mathematical arguments is it
possible to find an equation which under some prescribed restrictions will
reproduce the spectra of observables? Although it is widely believed that
both approaches are equivalent, already Dirac in his lectures on quantum
field theory [1] noticed (without much elaboration) that Schrodinger's
description of QM contains a lot of "dead wood" which can be safely disposed
altogether. According to Dirac \ "Heisenberg's picture of QM is good because
Heisenberg's equations of motion make sense". To our knowledge, Dirac's
comments were completely ignored, perhaps, because he had not provided
enough evidence making Heisenberg's description of QM superior to that of
Schrodinger's. In recent papers [2,3] we found examples supporting Dirac's
claims. From the point of view of combinatorics, there is not much
difference in description of QM, quantum field theory and string theory.
Therefore, in this paper we choose Heisenberg's point of view on string
theory using results of our recent works in which we re analyzed the
existing treatments connecting Veneziano (and Veneziano-like) amplitudes
with the respective string-theoretic models. As result, we were able to find
new tachyon-free models reproducing Veneziano (and Veneziano-like)
amplitudes. In this work results of our \ papers [4-7] which will be
respectively called as Part I, Part II, Part III \ and Part IV are \ further
developed to bring them in correspondence with those developed by other
authors. Without any changes in the \ already existing formalism, we were
able to connect our results with an impressive number of string-theoretic
models.

As is well known, all information in high energy physics is obtainable
through proper interpretation of the scattering data. It is believed that
for sufficiently high energies such data are well described by the
phenomenological Regge theory and can be conveniently summarized with help
of Chew-Frauthchi plots relating masses to spins (angular momenta), e.g. see
book by Collins[8]. Veneziano amplitudes are by design Regge-behaving. Both
Regge theory and Veneziano amplitudes emerged before major developments in
QCD took place in 70ies. Once these developments took place, naturally, it
was of interest to recover the Regge theory from QCD. Even though there are
many ways for doing so, to our knowledge, the problem is still not solved
completely. This is so for the following reasons.

Although the amount of data obtained by perturbative treatments of QCD is
quite impressive, e.g. read Ref.\textbf{[}9], these results are not helpful
for establishing the Regge-type behavior of QCD. \ Since such a behavior can
be easily established with help of variety of string models, the task lies
is connecting these models with QCD. Evidently, such a task is also equally
applicable to models reproduced in this work.

The connections between either QCD and spin chains or between strings and
spin chains were already discussed in literature for quite some time. Recent
paper by Dorey [10] contains may references \ listing these earlier results.
Subsequently, they had been replaced by those whose methods are based on
ADS/CFT correspondence. From the point of view of this correspondence, the
connections between strings and QCD also can be made through spin chains as
it is demonstrated in the seminal paper by Gubser, Klebanov and Polyakov,
Ref.[11]. Their ideas were developed in great detail in the paper by Minahan
and Zarembo, Ref.[12]. The spectrum of anomalous dimensions of operators in
the $\mathcal{N}=4,$ N$\rightarrow \infty $ supersymmetric Yang-Mills (Y-M)
model (described in terms of the excitation spectrum of the spin chain
model) is related to the string spectrum describing hadron masses. To
connect these facts with developments in this paper we mention papers by
Kruczenski [13], and Cotrone \textit{et al}.[14]. In both papers spin 1/2
XXX Heisenberg chain was used for description of the excitation spectrum.
Furthermore, in the paper by Cotrone\textit{\ et al }the explicit connection
with the hadron mass spectrum was made. Both papers use ADS/CFT
correspondence in their derivations. In this work, we reobtain these spin
chain results using combinatorial arguments following Heisenberg's
philosophy discussed in some detail in our recent work [2]. In view of this,
our derivation is not relying on use of ADS/CFT methods.

It should be noted though that the ADS/CFT is not the only method of
connecting QCD with strings. In 1981 't Hooft suggested [15] to reduce the
non Abelian QCD to Abelian Ginzburg-Landau (G-L) type theory. The rationale
for such an \textsl{Abelian} \textsl{reduction} can be traced back to the
work by Nambu [16]. In his work Nambu superimposed G-L theory with the
theory of Dirac monopoles to demonstrate quark confinement for mesons.
Incidentally, Veneziano amplitudes are suited the most for describing meson
resonances, e.g. see Ref.[8]. These are made of just two quarks: quark and
antiquark. Thus, if the existence of Abelian reduction would be considered
as proven, this then would be equivalent to the\ proof of quark confinement.
Recent numerical studies have provided convincing evidence supporting the
idea of quark confinement through monopole condensation, e.g. see
Refs.[17,18]. Since publication of 't Hooft's paper many theoretical
advancements were made, most notably by Cho, Ref.s[19,20], and Kondo, Ref.s
[21-23], whose work was motivated \ by that by Faddeev and his group.
Results of this group are summarized in\ the recent review by Faddeev,
Ref.[24]. From this reference it follows that most of efforts to date were
spent on description of the massless version of QCD. The excitation spectrum
of solitonic knotted-like structures (admitting interpretation in terms of
closed strings) provides the spectrum of glueball masses as demonstrated in
[22]. In the recent paper [25] Auckly, Kapitanski and Speight demonstrated
how Skyrme model can be obtained from Faddeev model. Since Skyrme model was
used for a long time for description of the baryon spectra, e.g. read
Ref.[26], and since already Nambu recognized usefulness of the Abelian
reduction for description of the meson spectra, it follows that the Abelian
reduction method is also capable of providing sufficient information about
QCD in the strong coupling regime.

Since our derivation is not relying on use of ADS/CFT methods, naturally,
the question arises: Can Heisenberg-style analysis of this work be useful in
deciding which of the methods: ADS/CFT or Abelian reduction is more likely
realized in Nature? In our companion paper [27] we provide evidence of
superiority of the Abelian reduction over ADS/CFT methods. To avoid
duplications, in this work we only develop connections between Veneziano
amplitudes, spin chains and various string models without making attempts to
relate explicitly these models with QCD.

The rest of this paper is organized as follows. In Sections 2-3 \ we provide
needed physical \ motivations and background. For this purpose, in Section 2
we replaced mathematically sophisticated derivation of Veneziano partition
function developed in [4-6] by a considerably simpler combinatorial
derivation of this function analogous to that discussed in [7]. As a
by-product of this effort, we were able to uncover connections with spin
chains already at this stage of our investigation. To strengthen this
connection, in Section 3 we demonstrate that \ the obtained Veneziano
partition function \ coincides with that for the Polychronakos-Frahm (P-F)
partition function for the ferromagnetic spin chain model [28]. In the limit
of infinitely long chains such a model coincides with spin 1/2 XXX
Heisenberg spin chain. This is explained in Section 4.\ In the same section
we discuss different paths aimed at establishing links between the P-F spin
chain and variety of string-theoretic models, including the most recent
ones. This is achieved by mapping \ the combinatorial and analytical
properties of the P-F spin chains into analogous properties of spin chains
used for description of the stochastic process known in literature as 
\textsl{asymptotic simple exclusion process} (ASEP). To make our
presentation self-contained, Appendix\ contains some basic information on
ASEP sufficient for understanding the results of the main text. In addition,
in the main text, we provide some information on the Kardar-Parisi-Zhang
(KPZ) and Edwards-Wilkinson (EW) equations which are just different well
defined macroscopic limits of the microscopic ASEP equations. This is done
with purpose of \ reproducing a variety of \ already known string-theoretic
models, including the most recent ones, and to relate them to each other.
Section 5 contains some discussion connecting the obtained results with that
presented in our companion paper [27].

\section{Combinatorics of Veneziano amplitudes and spin chains: qualitative
considerations}

In Part I, we noticed that the Veneziano condition for the 4-particle
amplitude given by 
\begin{equation}
\alpha (s)+\alpha (t)+\alpha (u)=-1,  \tag{1}
\end{equation}%
where $\alpha (s)$, $\alpha (t),\alpha (u)$ $\in \mathbf{Z}$, can be
rewritten in more mathematically suggestive form. To this purpose, following
[29], we need to consider additional homogenous equation of the type 
\begin{equation}
\alpha (s)m+\alpha (t)n+\alpha (u)l+k\cdot 1=0  \tag{2}
\end{equation}%
with $m,n,l,k$ being some integers. By adding this equation to (1) we obtain 
\begin{equation}
\alpha (s)\tilde{m}+\alpha (t)\tilde{n}+\alpha (u)\tilde{l}=\tilde{k} 
\tag{3a}
\end{equation}%
or, equivalently, 
\begin{equation}
n_{1}+n_{2}+n_{3}=\hat{N},  \tag{3b}
\end{equation}%
where all entries \textit{by design} are nonnegative integers. For the
multiparticle case this equation should be replaced by%
\begin{equation}
n_{0}+\cdot \cdot \cdot +n_{k}=N  \tag{4}
\end{equation}%
so that \textsl{combinatorially the task lies in finding all nonnegative
integer combinations} \textsl{of} $n_{0},...,n_{k}$ \textsl{producing} (4).
It should be noted that such a task makes sense as long as $N$ is assigned.
But the actual value of $N$ is \textit{not} \textit{fixed} and, hence, can
be chosen quite arbitrarily. Eq.(1) is a simple statement about the energy
-momentum conservation. Although the numerical entries in this equation can
be changed as we just explained, the actual physical values can be \
subsequently re obtained by the appropriate coordinate shift. \ The
arbitrariness in selecting $N$ reflects a kind of gauge freedom. As in other
gauge theories, we may try to fix the gauge by using some physical
considerations. These include, for example, an observation made in Part I
that the 4 particle amplitude is zero if any two entries into (1) are the
same. This \ fact prompts us to arrange the entries in (3b) in accordance
with their magnitude, i.e. $n_{1}\geq n_{2}\geq n_{3}.$ More generally, we
can write: $n_{0}\geq n_{1}\geq \cdot \cdot \cdot \geq n_{k}\geq 1\footnote{%
The last inequality: $n_{k}\geq 1,$ is chosen only for the sake of
comparison with the existing literature conventions, e.g. see Ref.[30\textbf{%
]}.}$.

Provided that (4) holds, we shall call such a sequence a \textit{partition }%
and shall denote it as\textit{\ }$\mathit{n\equiv }(n_{0},...,n_{k})$. If $n$
is a partition of $N$, then we shall write $n\vdash N$. It is well known
[30,31] that there is one- to -one correspondence between the Young diagrams
and partitions. We would like to use this fact in order to design a
partition function capable of reproducing Veneziano (and Veneziano-like)
amplitudes. Clearly, such a partition function should also make physical
sense. Hence, we would like to provide some qualitative arguments aimed at
convincing our readers that such a partition function does exist and is
physically sensible.

We begin with observation that there is one- to- one correspondence between
the Young tableaux and directed random walks\footnote{%
Furthermore, it is possible to map bijectively such type of random walk back
into Young diagram with only two rows, e.g. read [32], page 5. This allows
us to make a connection with spin chains at once. In this work we are not
going to use this route to spin chains in view of the simplicity of \
alternative approaches discussed in this section.}. It is useful to recall
details of this correspondence now. To this purpose we need to consider a
square lattice and to place on it the Young diagram associated with some
particular partition.

Let us choose some $\tilde{n}\times \tilde{m}$ rectangle\footnote{%
Parameters $\tilde{n}$ and $\tilde{m}$ will be specified shortly below.} so
that the Young diagram occupies the left part of this rectangle. We choose
the upper left vertex of the rectangle as the origin of the $xy$ coordinate
system whose $y$ axis (South direction) is directed downwards and $x$ axis
is directed Eastwards. Then, the South-East boundary of the Young diagram
can be interpreted as directed (that is without self intersections) random
walk which begins at $(0,-\tilde{m})$ and ends at $(\tilde{n},0).$
Evidently, such a walk completely determines the diagram. The walk can be
described by a sequence of 0's and 1's.\ Say, $0$ for the $x-$ step move and
1 for the $y-$ step move. The totality $\mathcal{N}$ of Young diagrams which
can be placed into such a rectangle is in one-to-one correspondence with the
number of arrangements of 0's and 1's whose total number is $\tilde{m}+%
\tilde{n}$. Recalling the Fermi statistics, the number $\mathcal{N}$ can be
easily calculated and is given by $\mathcal{N}=(m+n)!/m!n!\footnote{%
We have suppressed the tildas for $n$ and $m$ in this expression since these
parameters are going to be redefined below anyway.}$. It can be represented
in two equivalent ways%
\begin{eqnarray}
(m+n)!/m!n! &=&\frac{(n+1)(n+2)\cdot \cdot \cdot (n+m)}{m!}\equiv \left( 
\begin{array}{c}
n+m \\ 
m%
\end{array}%
\right)  \notag \\
&=&\frac{(m+1)(m+2)\cdot \cdot \cdot (n+m)}{n!}\equiv \left( 
\begin{array}{c}
m+n \\ 
n%
\end{array}%
\right) .  \TCItag{5}
\end{eqnarray}

Let now $p(N;k,m)$ be the number of partitions of $N$ into $\leq k$ \
nonnegative parts, each not larger than $m$. Consider the generating
function of the following type 
\begin{equation}
\mathcal{F}(k,m\mid q)=\dsum\limits_{N=0}^{S}p(N;k,m)q^{N}  \tag{6}
\end{equation}%
where the upper limit $S$\ will be determined shortly below. It is shown in
Refs.[30,31\textbf{]} that $\mathcal{F}(k,m\mid q)=\left[ 
\begin{array}{c}
k+m \\ 
m%
\end{array}%
\right] _{q}\equiv \left[ 
\begin{array}{c}
k+m \\ 
k%
\end{array}%
\right] _{q}$ where, for instance,$\left[ 
\begin{array}{c}
k+m \\ 
m%
\end{array}%
\right] _{q=1}=\left( 
\begin{array}{c}
k+m \\ 
m%
\end{array}%
\right) \footnote{%
On page 15 of the book by Stanley [31], one can find that the number of
solutions $N(n,k)$ in \textit{positive} integers to $y_{1}+...+y_{k}=n+k$ is
given by $\left( 
\begin{array}{c}
n+k-1 \\ 
k-1%
\end{array}%
\right) $ while the number of solutions in \textit{nonnegative} integers to $%
x_{1}+...+x_{k}=n$ is $\left( 
\begin{array}{c}
n+k \\ 
k%
\end{array}%
\right) .$ Careful reading of Page 15 indicates however that the last number
refers to solution in nonnegative integers of the equation $%
x_{0}+...+x_{k}=n $. This fact was used essentially in (1.21) of Part I.}.$
From this result it should be clear that the expression $\left[ 
\begin{array}{c}
k+m \\ 
m%
\end{array}%
\right] _{q}$ is the $q-$analog of the binomial coefficient $\left( 
\begin{array}{c}
k+m \\ 
m%
\end{array}%
\right) .$ In literature [30,31] this $q-$ analog is known as the \textit{%
Gaussian} coefficient. Explicitly, it is defined as%
\begin{equation}
\left[ 
\begin{array}{c}
a \\ 
b%
\end{array}%
\right] _{q}=\frac{(q^{a}-1)(q^{a-1}-1)\cdot \cdot \cdot (q^{a-b+1}-1)}{%
(q^{b}-1)(q^{b-1}-1)\cdot \cdot \cdot (q-1)}  \tag{7}
\end{equation}%
for some nonegative integers $a$ and $b$. From this definition we anticipate
that the sum defining generating function $\mathcal{F}(k,m\mid q)$ in (6)
should have only \textit{finite} number of terms. Eq.(7) allows easy
determination of the upper limit $S$ in the sum (6). It is given by $km$.
This is just the area of the $k\times m$ rectangle. In view of the
definition of $p(N;k,m)$, the number $m=N-k$. Using this fact (6) can be
rewritten as: $\mathcal{F}(N,k\mid q)=\left[ 
\begin{array}{c}
N \\ 
k%
\end{array}%
\right] _{q}.$This expression happens to be the Poincare$^{\prime }$
polynomial for the Grassmannian $Gr(m,k)$ of the complex vector space 
\textbf{C}$^{N}$of dimension $N$ as can be seen from page 292 of the book by
Bott and Tu, [33]\footnote{%
To make a comparison it is sufficient to replace parameters $t^{2}$ and $n$
in \ Bott and Tu book by $q$ and $N.$}. From this (topological) point of
view the numerical coefficients, i.e. $p(N;k,m),$ in the $q$ expansion of
(6) should be interpreted as Betti numbers of this Grassmannian. They can be
determined recursively using the following property of the Gaussian
coefficients [31], page 26,%
\begin{equation}
\left[ 
\begin{array}{c}
n+1 \\ 
k+1%
\end{array}%
\right] _{q}=\left[ 
\begin{array}{c}
n \\ 
k+1%
\end{array}%
\right] _{q}+q^{n-k}\left[ 
\begin{array}{c}
n \\ 
k%
\end{array}%
\right] _{q}  \tag{8}
\end{equation}%
and taking into account that $\left[ 
\begin{array}{c}
n \\ 
0%
\end{array}%
\right] _{q}=1.$ We refer our readers to Part II for mathematical proof that 
$\mathcal{F}(N,k\mid q)$ is indeed the Poincare$^{\prime }$ polynomial for
the complex Grassmannian. With this fact proven, we notice that, due to
relation $m=N-k,$ it is sometimes more convenient \ for us to use the
parameters $m$ and $k$ rather than $N$ and $k$. With such a replacement we
obtain: 
\begin{eqnarray}
\mathcal{F}(k,m &\mid &q)=\left[ 
\begin{array}{c}
k+m \\ 
k%
\end{array}%
\right] _{q}=\frac{(q^{k+m}-1)(q^{k+m-1}-1)\cdot \cdot \cdot (q^{m+1}-1)}{%
(q^{k}-1)(q^{k-1}-1)\cdot \cdot \cdot (q-1)}  \notag \\
&=&\dprod\limits_{i=1}^{k}\frac{1-q^{m+i}}{1-q^{i}}.  \TCItag{9}
\end{eqnarray}%
This result is of central importance for this work. In Part II a
considerably more sophisticated mathematical apparatus was used to obtain it
(e.g. see equation (6.10) of this reference and arguments leading to it).\ 

In the limit $q\rightarrow 1$ Eq.(9) reduces to $\mathcal{N}$ as required.
To make connections with results known in physics literature we need to re
scale $q^{\prime }s$ in (9), e.g. let $q=t^{\frac{1}{i}}.$ Substitution of
such an expression back into (9) and taking the limit $t\rightarrow 1$ again
produces $\mathcal{N}$ in view of (5). This time, however, we can accomplish
more. By noticing that in (4) the actual value of $N$ deliberately is not
yet fixed and taking into account that $m=N-k$ we can fix $N$ by fixing $m$.
Specifically, we would like to choose $m=1\cdot 2\cdot 3\cdot \cdot \cdot k$
and with such a choice we would like\ to consider a particular term in the
product (9), e.g.%
\begin{equation}
S(i)=\frac{1-t^{1+\frac{m}{i}}}{1-t}.  \tag{10}
\end{equation}%
In view of our "gauge fixing" the ratio $m/i$ is a positive integer by
design. This means that we are having a geometric progression. Indeed, if we
rescale $t$ again : $t\rightarrow t^{2},$ we then obtain:%
\begin{equation}
S(i)=1+t^{2}+\cdot \cdot \cdot +t^{2\hat{m}}  \tag{11}
\end{equation}%
with $\hat{m}=\frac{m}{i}.$ Written in such a form the \ above sum is just
the Poincare$^{\prime }$ polynomial for the complex projective space \textbf{%
CP}$^{\hat{m}}.$ This can be seen by comparing pages 177 and 269 of the book
by Bott and Tu [33]. Hence, at least for some $m$'s, \textit{the Poincare}$%
^{\prime }$\textit{\ polynomial for the Grassmannian in just the product of} 
\textit{the Poincare}$^{\prime }$\textit{\ polynomials for the complex
projective spaces of known dimensionalities}. For $m$ just chosen, in the
limit $t\rightarrow 1,$ we reobtain back the number $\mathcal{N}$ as
required. This physically motivating process of gauge fixing just described
can be replaced by more rigorous mathematical arguments. The recursion
relation (8) indicates that this is possible. The \ mathematical details
leading to factorization which we just described can be found, for instance,
in the Ch-3 of lecture notes by Schwartz [34]. The relevant physics emerges
by noticing that the partition function $Z(J)$ for the particle with spin $J$
is given by [35] 
\begin{eqnarray}
Z(J) &=&tr(e^{-\beta H(\sigma )})=e^{cJ}+e^{c(J-1)}+\cdot \cdot \cdot
+e^{-cJ}  \notag \\
&=&e^{cJ}(1+e^{-c}+e^{-2c}+\cdot \cdot \cdot +e^{-2cJ}),  \TCItag{12}
\end{eqnarray}%
where $c$ is known constant. Evidently, up to a constant, $Z(J)\simeq S(i).$
Since\ mathematically the result (12) is the Weyl character formula, this
observation brings the classical group theory into our discussion. More
importantly, because the partition function for the particle with spin $J$
can be written in the language of N=2 supersymmetric quantum mechanical model%
\footnote{%
We hope that no confusion is made about the meaning of N in the present case.%
}, as demonstrated by Stone [35] and others [36], the connection between the
supersymmetry and the classical group theory is evident. \ It was developed
in Parts III and IV.

In view of arguments presented above, the Poincare$^{\prime }$ polynomial
for the Grassmannian can be interpreted as a partition function for some
kind of a spin chain made of \ apparently independent spins of various
magnitudes\footnote{%
In such a context it can be vaguely considered as a variation on the theme
of the Polyakov rigid string (Grassmann $\sigma $ model, Ref.[37], pages
283-287), except that now it is \textit{exactly solvable} in the qualitative
context \ just described and, below, in the mathematically rigorous context.}%
. These qualitative arguments we would like to make more mathematically and
physically rigorous. The first step towards this goal is made in the next
section.

\section{Connection with the Polychronakos-Frahm spin chain model}

\bigskip

The Polychronakos-Frahm (P-F) spin chain model \ was originally proposed by
Polychronakos and described in detail in [28]. Frahm [38] \ motivated by the
results of Polychronakos made additional progress in elucidating the
spectrum and thermodynamic properties of this model so that it had become
known as the P-F model. Subsequently, many other researchers have
contributed to our understanding of this exactly integrable spin chain
model. Since this paper is not a review, we shall quote only works on P-F
model which are of immediate relevance.

Following [28], we begin with some description of the P-F model. Let $\sigma
_{i}^{a}$ ($a=1,2,...,n^{2}-1)$ be $SU(n)$ spin operator of i-th particle
and let the operator$\ \sigma _{ij}$ be responsible for a spin exchange
between particles $i$ and $j,$ i.e.%
\begin{equation}
\sigma _{ij}=\frac{1}{n}+\tsum\limits_{a}\sigma _{i}^{a}\sigma _{j}^{a}. 
\tag{13}
\end{equation}%
In terms of these definitions, the Calogero-type model Hamiltonian can be
written as [39,40] 
\begin{equation}
\mathcal{H}=\frac{1}{2}\tsum\limits_{i}(p_{i}^{2}+\omega
^{2}x_{i}^{2})+\tsum\limits_{i<j}\frac{l(l-\sigma _{ij})}{\left(
x_{i}-x_{j}\right) ^{2}},  \tag{14}
\end{equation}%
where $l$ is some parameter. The P-F model is obtained from the above model
in the limit $l\rightarrow \pm \infty $ . Upon proper rescaling of $\mathcal{%
H}$ in (14), in this limit one obtains%
\begin{equation}
\mathcal{H}_{\mathcal{P-F}}=-sign(l)\tsum\limits_{i<j}\frac{\sigma _{ij}}{%
\left( x_{i}-x_{j}\right) ^{2}},  \tag{15}
\end{equation}%
where the coordinate $x_{i}$ minimizes the potential for the rescaled
Calogero model\footnote{%
The Calogero model is obtainable from the Hamiltonian (14) if one replaces
the spin exchange operator $\sigma _{ij}$ by 1. Since we are interested in
the large $l$ limit, one can replace the factor $l(l-1)$ by $l^{2}$ in the
interaction term.}, that is%
\begin{equation}
\omega ^{2}x_{i}^{{}}=\tsum\limits_{i<j}\frac{2}{\left( x_{i}-x_{j}\right)
^{3}}.  \tag{16}
\end{equation}%
It should be noted that $\mathcal{H}_{\mathcal{P-F}}$ is well defined
without such a minimization, that is for arbitrary real parameters $x_{i}$.
This fact is explained in detail in our companion paper[27].\ In the large $%
l $ limit the spectrum of $\mathcal{H}$ is decomposable as 
\begin{equation}
E=E_{\mathcal{C}}+lE_{\mathcal{P-F}},  \tag{17}
\end{equation}%
where $E_{C}$ is the spectrum of the spinless Calogero model while $E_{%
\mathcal{P-F}}$ is the spectrum of the P-F model. In view of such a
decomposition, the partition function for the Hamiltonian $\mathcal{H}$ at
temperature $T$ can be written as a product: Z$_{\mathcal{H}}(T)=$Z$_{%
\mathcal{C}}(T)$Z$_{\mathcal{P-F}}(T/l)$. From here, one formally obtains
the result: 
\begin{equation}
Z_{\mathcal{P-F}}(T)=\lim_{l\rightarrow \infty }\frac{Z_{\mathcal{H}}(lT)}{%
Z_{\mathcal{C}}(T)}.  \tag{18}
\end{equation}%
It implies that the spectrum of the P-F spin chain can be obtained if both
the total and \ the Calogero partition functions can be calculated. In [28]
Polychronakos argued that $Z_{\mathcal{C}}(T)$ is essentially a partition
function of $\mathit{N}$ noninteracting harmonic oscillators. Thus, we
obtain 
\begin{equation}
Z_{\mathcal{C}}(N;T)=\tprod\limits_{i=1}^{N}\frac{1}{1-q^{i}},\text{ }q=\exp
(-\beta ),\beta =\left( k_{B}T\right) ^{-1}.  \tag{19}
\end{equation}%
Furthermore, the partition function $Z_{\mathcal{H}}(T)$ according to
Polychronakos can be obtained using $Z_{\mathcal{C}}(N;T)$ as follows.
Consider the grand partition function of the type%
\begin{equation}
\Xi =\tsum\limits_{N=0}^{\infty }Z_{n}(N;T)y^{N}\equiv \left(
\tsum\limits_{L=0}^{\infty }Z_{\mathcal{C}}(L;T)y^{L}\right) ^{n}  \tag{20}
\end{equation}%
where $n$ is the number of flavors\footnote{%
That is $n$ the same number as $n$ in $SU(n).$}. \ Using this definition we
obtain%
\begin{equation}
Z_{n}(N;T)=\sum_{\Sigma _{i}k_{i}=N}\prod\limits_{i=1}^{n}Z_{\mathcal{C}%
}(k_{i};T).  \tag{21}
\end{equation}%
Next, Polychronakos identifies $Z_{n}(N;T)$ with Z$_{\mathcal{H}}(T)$. Then,
with help of (18) the partition function $Z_{\mathcal{P-F}}(T)$ is obtained
straightforwardly as 
\begin{equation}
Z_{\mathcal{P-F}}(N;T)=\sum_{\Sigma _{i}k_{i}=N}\frac{\tprod%
\limits_{i=1}^{N}(1-q^{i})}{\prod\limits_{i=1}^{n}\prod%
\limits_{r=1}^{k_{i}^{{}}}(1-q^{r})}.  \tag{22}
\end{equation}%
Consider this result for a special case: $n=2$. It is convenient to evaluate
the ratio first before calculating the sum. Thus, we obtain%
\begin{equation}
\frac{\tprod\limits_{i=1}^{N}(1-q^{i})}{\prod\limits_{i=1}^{2}\prod%
\limits_{r=1}^{k_{i}^{{}}}(1-q^{r})}=\frac{(1-q)\cdot \cdot \cdot (1-q^{N})}{%
(1-q)\cdot \cdot \cdot (1-q^{k})(1-q)\cdot \cdot \cdot (1-q^{N-k})}\equiv 
\mathcal{F}(N,k\mid q),  \tag{23}
\end{equation}%
where the Poincare$^{\prime }$ polynomial $\mathcal{F}(N,k\mid q)$ for the
Grassmanian of the complex vector space \textbf{C}$^{N}$ of dimension $N$ \
was obtained in the previous section. Indeed (23) can be trivially brought
into the same form as that given in our Eq.(9) using the relation $m+k=N$.
To bring Eq.(9) in correspondence with equation (4.1) of Polychronakos [28],
we use the second equality given in \ Eq.(9) \ in which we make a
substitution: $m=N-k.$ After this replacement, Eq.(22) acquires the form%
\begin{equation}
Z_{\mathcal{P-F}}^{f}(N;T)=\sum\limits_{k=0}^{N}\prod\limits_{i=0}^{k}\frac{%
1-q^{N-i+1}}{1-q^{i}}  \tag{24}
\end{equation}%
coinciding with \ Eq.(4.1) by Polychronakos. This equation corresponds to
the ferromagnetic version of the P-F spin chain model. To obtain the
antiferromagnetic version of the model requires us only to replace $q$ by $%
q^{-1}$ in (24) and to multiply the whole r.h.s. by some known power of $q$.
Since this factor will not affect thermodynamics, following Frahm [38], we
shall ignore it. As result, we obtain 
\begin{equation}
Z_{\mathcal{P-F}}^{af}(N;T)=\sum\limits_{k=0}^{N}q^{\left( N/2-k\right)
^{2}}\prod\limits_{i=0}^{k}\frac{1-q^{N-i+1}}{1-q^{i}},  \tag{25}
\end{equation}%
in accord with Frahm's Eq.(21). \ This result is analyzed further in the
next section playing the central role for this paper.

\section{Connections with WZNW model and XXX \ s=1/2 \ Heisenberg
antiferromagnetic \ spin chain}

\subsection{General remarks}

\bigskip To establish these connections we follow work by Hikami [41]. For
this purpose, we introduce the notation%
\begin{equation}
\left( q\right) _{n}=\prod\limits_{i=1}^{n}(1-q^{i})  \tag{26}
\end{equation}%
allowing us to rewrite (25) in the equivalent form%
\begin{equation}
Z_{\mathcal{P-F}}^{af}(N;T)=\sum\limits_{k=0}^{N}q^{\left( N/2-k\right)
^{2}}\prod\limits_{i=0}^{k}\frac{1-q^{N-i+1}}{1-q^{i}}=\sum%
\limits_{k=0}^{N}q^{\left( N/2-k\right) ^{2}}\frac{\left( q\right) _{N}}{%
\left( q\right) _{k}\left( q\right) _{N-k}}.  \tag{27}
\end{equation}%
Consider now the limiting case ($N\rightarrow \infty )$ of the obtained
expression. \ For this purpose we need to take into account that 
\begin{equation}
\lim_{N\rightarrow \infty }\left[ 
\begin{array}{c}
N \\ 
k%
\end{array}%
\right] _{q}=\frac{1}{\left( q\right) _{k}}.  \tag{28}
\end{equation}%
To use this asymptotic result in (27) it is convenient to consider
separately the cases of $N$ being even and odd. For instance, if $N$ is
even, we can write: $N=2m.$ In such a case we can introduce new summation
variables: $l=k-m$ and/or $l=m-k.$ Then, in the limit $N\rightarrow \infty $
(that is $\mathit{m}\rightarrow \infty )$ we obtain asymptotically%
\begin{equation}
Z_{\mathcal{P-F}}^{af}(\infty ;T)=\frac{1}{\left( q\right) _{\infty }}%
\sum\limits_{i=-\infty }^{\infty }q^{i^{2}}.  \tag{29a}
\end{equation}%
in accord with [41]. Analogously, if $N=2m+1$, we obtain instead 
\begin{equation}
Z_{\mathcal{P-F}}^{af}(\infty ;T)=\frac{1}{\left( q\right) _{\infty }}%
\sum\limits_{i=-\infty }^{\infty }q^{\left( i+\frac{1}{2}\right) ^{2}}. 
\tag{29b}
\end{equation}%
According to Melzer [42] and Kedem, McCoy and Melzer [43], the obtained
partition functions coincide with the Virasoro characters for SU$_{1}$(2)
WZNW \ model describing the conformal limit of the XXX (s=1/2)
antiferromagnetic spin chain [44]. Even though equations (29a) and (29b)
provide the final result, they do not reveal their physical content. This
task was accomplished in part in the same papers where connection with the
excitation spectrum of the XXX antiferromagnetic chain \ was made. To avoid
repetitions, \ below we arrive at these conclusions using different
arguments. By doing so many new and unexpected connections \ with other \
stochastic models will be uncovered.

\subsection{Method of generating functions and q-deformed harmonic oscillator%
}

We begin with definitions. In view of \ equations (9),(24) and (27), we
would like to introduce the Galois number $G_{N}$ via 
\begin{equation}
G_{N}=\sum\limits_{k=0}^{N}\left[ 
\begin{array}{c}
N \\ 
k%
\end{array}%
\right] _{q}.  \tag{30}
\end{equation}%
This number can be calculated recursively as it was shown by Goldman and
Rota [45] with the result%
\begin{equation}
G_{N+1}=2G_{N}+\left( q^{N}-1\right) G_{N-1}.  \tag{31}
\end{equation}%
Alternative proof was given by Kac and Cheung [46]. To calculate $G_{N}$ we
have to take into account that $G_{0}=1$ and $G_{1}=2.$ These results can be
used as a reference when one attempts to calculate the related Rogers-Szego\
(R-S) polynomial $H_{N}(t)$ defined as [47]%
\begin{equation}
H_{N}(t;q):=\sum\limits_{k=0}^{N}\left[ 
\begin{array}{c}
N \\ 
k%
\end{array}%
\right] _{q}t^{k}  \tag{32}
\end{equation}%
so that $H_{N}(1)=G_{N}\footnote{%
For brevity, unless needed explicitly, we shall suppress the argument $q$ in 
$H_{N}(t;q).$}.$ Using Ref.[46] once again, we find that $H_{N}(t)$ obeys
the following recursion relation%
\begin{equation}
H_{N+1}(t)=(1+t)H_{N}(t)+\left( q^{N}-1\right) tH_{N-1}(t)  \tag{33}
\end{equation}%
which for $t=1$ coincides with (31) as required. The above recursion
relation is supplemented with initial conditions. These are : $H_{0}=1$ and $%
H_{1}=1+t$.

At this point we would like to remind our readers that for $t=1$ according
to (24),(30) and (32) we obtain: $Z_{\mathcal{P-F}}^{f}(N;T)=G_{N}$. Hence,
by calculating $H_{N}(t)$ we also shall obtain the partition function for
the P-F chain.

To proceed with such calculations, we follow Ref.[48]. In particular, we
consider first the auxiliary recursion relation for Hermite polynomials:%
\begin{equation}
H_{n+1}(x)=2xH_{n}(x)-2nH_{n-1}(x)  \tag{34a}
\end{equation}%
supplemented by the differential relation 
\begin{equation}
\frac{d}{dx}H_{n}(x)=2nH_{n-1}(x)  \tag{34b}
\end{equation}%
which, in view of (34a), can be conveniently rewritten as 
\begin{equation}
H_{n+1}(x)=(2x-\frac{d}{dx})H_{n}(x).  \tag{34c}
\end{equation}%
This observation prompts us to introduce the raising operator $R=2x-\dfrac{d%
}{dx}$ so that we obtain: 
\begin{equation}
R^{n}H_{0}(x)=H_{n}(x).  \tag{35}
\end{equation}%
The lowering operator can be now easily obtained again using (34). \ Indeed,
we obtain 
\begin{equation}
\frac{1}{2}\frac{d}{dx}H_{n}(x)\equiv LH_{n}(x)=nH_{n-1}(x)  \tag{36}
\end{equation}%
so that $[L,R]=1$ as required. \ Based on this, the number operator $N$ can
be obtained as $N=RL$ so that $NH_{n}(x)=nH_{n}(x)$ or, explicitly, using
provided definitions, we obtain: 
\begin{equation}
(\frac{d^{2}}{dx^{2}}-2x\frac{d}{dx}+2n)H_{n}(x)=0.  \tag{37}
\end{equation}%
Evidently, we can write: $R\mid n>=\mid n+1>,$ $L\mid n>=\mid n-1>$ and , $%
<m\mid n>=n!\delta _{mn}$ as usual.

We would like now to transfer all these results to our main object of
interest- the recursion relation (33). To this purpose, we introduce the
difference operator $\Delta $ via 
\begin{equation}
\Delta H_{N}(t):=H_{N}(t)-H_{N}(qt).  \tag{38}
\end{equation}%
Using (32) we obtain now 
\begin{equation}
\Delta H_{N}(t)=(1-q^{N})tH_{N-1}(t),  \tag{39}
\end{equation}%
where we took into account that 
\begin{equation}
\left[ 
\begin{array}{c}
N \\ 
k%
\end{array}%
\right] _{q}=\left[ 
\begin{array}{c}
N \\ 
N-k%
\end{array}%
\right] _{q}.  \tag{40}
\end{equation}%
Using this result in (33) we obtain at once 
\begin{equation}
H_{N+1}(t)=[(1+t)-\Delta ]H_{N}(t).  \tag{41}
\end{equation}%
This, again, can be looked upon as a definition of a raising operator so
that we can formally rewrite (41) as 
\begin{equation}
\mathcal{R}H_{N}(t)=H_{N+1}(t).  \tag{42}
\end{equation}%
The lowering operator can be defined now as 
\begin{equation}
\mathcal{L}:=\frac{1}{x}\Delta  \tag{43}
\end{equation}%
so that 
\begin{equation}
\mathcal{L}H_{N}(t)=(1-q^{N})H_{N-1}(t).  \tag{44}
\end{equation}%
The action of the number operator $\mathcal{N}=\mathcal{RL}$ is \ now
straightforward, i.e.%
\begin{equation}
\mathcal{N}H_{N}(t)=(1-q^{N})H_{N}(t).  \tag{45}
\end{equation}%
Following Kac and Cheung [\textbf{32}] we introduce the $q-$derivative via 
\begin{equation}
D_{q}f(x):=\frac{f(qx)-f(x)}{x(q-1)}.  \tag{46}
\end{equation}%
By combining this result with (38) we obtain, 
\begin{equation}
D_{q}f(x)=\frac{\Delta f(x)}{x(1-q)}.  \tag{47}
\end{equation}%
This allows us to rewrite the raising and lowering operators in terms of $q-$%
derivatives. Specifically, we obtain%
\begin{equation}
\mathcal{\tilde{R}}\text{ :}\mathcal{=}(1+t)-(1-q)tD_{q},  \tag{48}
\end{equation}%
and 
\begin{equation}
\mathcal{\tilde{L}}:=D_{q}.  \tag{49}
\end{equation}%
While for the raising operator rewritten in such a way Eq.(42) still holds,
for the lowering operator $\mathcal{\tilde{L}}$ we now obtain:%
\begin{equation}
\mathcal{\tilde{L}}H_{N}(t)=\frac{1-q^{N}}{1-q}H_{N-1}(t)\equiv \lbrack
N]H_{N-1}(t).  \tag{50}
\end{equation}%
The number operator $\mathcal{N}_{q}$ is acting in this case as 
\begin{equation}
\mathcal{N}_{q}H_{N}(t)=[N]H_{N}(t).  \tag{51}
\end{equation}%
We would like to connect these results with those available in literature on 
$q-$deformed harmonic oscillator. Following Chaichan et al [49], we notice
that the undeformed oscillator algebra is given in terms of the following
commutation relations 
\begin{equation}
aa^{+}-a^{+}a=1  \tag{52a}
\end{equation}%
\begin{equation}
\lbrack N,a]=-a  \tag{52b}
\end{equation}%
and%
\begin{equation}
\lbrack N,a^{+}]=a^{+}.  \tag{52c}
\end{equation}%
In these relations it is not assumed a priori that $N=a^{+}a$ and,
therefore, this algebra is formally different from the traditionally used $%
[a,a^{+}]=1$ for the harmonic oscillator. This observation allows us to
introduce the central element $Z=N-a^{+}a$ which is zero for the standard
oscillator algebra. The deformed oscillator algebra can be obtained now
using Eq.s(52) in which one should replace (52a) by 
\begin{equation}
aa^{+}-qa^{+}a=1.  \tag{52d}
\end{equation}%
as it is done in [50]. Consider now the combination $K:=$ $\mathcal{LR}$-$q%
\mathcal{N}$ acting on $H_{N}$ using previously introduced definitions.
Simple calculation produces an operator identity: $\mathcal{LR}-q\mathcal{N=}%
1,$ so that we can formally make a provisional identification : $\mathcal{%
L\rightarrow }a$ and $\mathcal{R\rightarrow }a^{+}.$ To proceed, we need to
demonstrate that with such an identification equations (52 b,c) hold as
well. For this to happen, we should \ properly normalize our wave function
in accord with known procedure for the harmonic oscillator where we have to
use $\mid n>=\dfrac{1}{\sqrt{n!}}\left( a^{+}\right) ^{n}\mid 0>$. In the
present case, we have to use $\mid N>=\dfrac{1}{\sqrt{[N]!}}\left( \mathcal{R%
}\right) ^{n}\mid 0>$ as the basis wavefunction while making an
identification: $\mid 0>=$ $H_{0}(t).$ The eigenvalue equation (51), when
written explicitly, acquires the following form:%
\begin{equation}
\lbrack tD_{q}^{2}-\frac{1+t}{1-q}D_{q}+\frac{[N]}{1-q}]H_{N}(t)=0.  \tag{53}
\end{equation}

\subsection{The limit $q\rightarrow 1^{\pm }$ and emergence of the
Stieltjes-Wigert polynomials}

\bigskip Obtained results need further refinements for the following
reasons. Although the recursion relations (33), (34) look similar, in the
limit $q\rightarrow 1^{\pm }$ equation (33) is not transformed into (34).
Accordingly, (53) is not converted into equation for Hermite polynomials
known for harmonic oscillator. Fortunately, the situation can be repaired in
view of recent paper by Karabulut [51] who spotted and corrected some error
in the influential earlier paper by Macfarlane [52]. Following the logic of
\ these papers, we define the translation operator $T(s)$ as $T(s):=e^{s%
\frac{\partial }{\partial x}}$. Using this definition, the creation $a^{\dag
}$ and annihilation $a$ operators are defined as follows%
\begin{equation}
a^{\dag }=\frac{1}{\sqrt{1-q}}[q^{x+\frac{1}{4}}-T^{\frac{-1}{2}}(s)]T^{%
\frac{-1}{2}}(s)  \tag{54a}
\end{equation}%
where $T^{\frac{1}{2}}(s)=e^{\frac{s}{2}\frac{\partial }{\partial x}}$ and,
accordingly, 
\begin{equation}
a=\frac{1}{\sqrt{1-q}}T^{\frac{1}{2}}(s)[q^{x+\frac{1}{4}}-T^{\frac{1}{2}%
}(s)].  \tag{54b}
\end{equation}%
Under such conditions, the inner product is defined in the standard way,
that is%
\begin{equation}
(f,g)=\int\limits_{-\infty }^{\infty }f^{\ast }(x)g(x)dx  \tag{55}
\end{equation}%
so that $(q^{x})^{\dag }=q^{x}$ and $(\partial /\partial x)^{\dag
}=-(\partial /\partial x)$ thus making the operator $a^{\dag }$ to be a
conjugate of $a$ in a usual way. The creation-annihilation operators just
defined satisfy commutation relation (52d). At the same time, the
combination $a^{\dag }a\ $\ while acting on the wave functions $\Psi _{n%
\text{ }}$\ (to be defined below)\ \ produces equation similar to (51), that
is 
\begin{equation}
a^{\dag }a\Psi _{n\text{ }}=[n]\Psi _{n\text{ }}\equiv \lambda _{n}\Psi _{n%
\text{ }}.  \tag{56}
\end{equation}%
\ Furthermore, it can be shown, that \ \ \ 
\begin{equation}
a\Psi _{n\text{ }}=\sqrt{\lambda _{n}}\Psi _{n-1\text{ }}\text{ \ \ and \ }%
a^{\dag }\Psi _{n\text{ }}=\sqrt{\lambda _{n+1}}\Psi _{n+1\text{ }}  \tag{57}
\end{equation}%
\ in accord with previously obtained results. Next, we would like to obtain
the wave function $\Psi _{n\text{ }}$ explicitly. To this purpose we start
with the ground state $a\Psi _{0\text{ }}=0$ and use (54b) to get (for $%
s=1/2 $)\footnote{%
The rationale for choosing $s=1/2$ is explained in the same reference.} the
following result 
\begin{equation}
\Psi _{0\text{ }}(x+\frac{1}{2})=q^{\frac{1}{4}+x}\Psi _{0\text{ }}(x). 
\tag{58}
\end{equation}%
\ \ Let $w(x)$ be some yet unknown function. Then, it is appropriate to look
for solution of (58) in the form%
\begin{equation}
\Psi _{0\text{ }}(x)=const\text{ }\cdot w(x)q^{x^{2}},  \tag{59a}
\end{equation}%
provided that the function $w(x)$ is periodic: \ $w(x)=w(x+1/2).$ The
normalized ground state function acquires then the following look%
\begin{equation}
\Psi _{0\text{ }}(x)=\alpha _{w}w(x)q^{x^{2}},  \tag{59b}
\end{equation}%
where the constant $\alpha _{w}$ is given by 
\begin{equation}
\alpha _{w}=\left( \int\limits_{-\infty }^{\infty }dx\left\vert
q^{x^{2}}w(x)\right\vert ^{2}\right) ^{-\frac{1}{2}}.  \tag{59c}
\end{equation}%
Using this result,\ \ $\Psi _{n\text{ }}$ can be constructed in a standard
way through use of the raising operators. There is, however, a faster way to
obtain the desired result.\ \ To this purpose, in view of (59b), suppose
that $\Psi _{n\text{ }}(x)$ can be decomposed as follows 
\begin{equation}
\Psi _{n\text{ }}(x)=\frac{\alpha _{w}w(x)}{\sqrt{(q,q)_{n}}}%
\sum\limits_{k=0}^{\infty }C_{k}^{n}(q)(-1)^{k}q^{\left( n-k\right)
/2}q^{(x-k)^{2}},  \tag{60}
\end{equation}%
\ \ \ where\ $(q,q)_{n}=(1-q)(1-q^{2})\cdot \cdot \cdot (1-q^{n})$ and $%
C_{k}^{n}(q)$ is to be determined. To do so, by applying the operator \ $%
a^{\dag }/$\ $\sqrt{\lambda _{n+1}}$ to (60)\ and taking into account that \ 
$T^{-\frac{1}{2}}$\ $w(x)=w(x-1/2)=w(x)$ (in \ view of the periodicity of $%
w(x)$) we end up with the recursion relation for $C_{k}^{n}(q)$:%
\begin{equation}
C_{k}^{n+1}(q)=q^{k}C_{k}^{n}(q)+C_{k-1}^{n}(q).  \tag{61}
\end{equation}%
\ This relation should be compared with that given by (8). Andrews [47],
page 35, demonstrated that (8) and (61) are equivalent. That is,%
\begin{equation}
C_{k}^{n}(q)=\left[ 
\begin{array}{c}
n \\ 
k%
\end{array}%
\right] _{q}.  \tag{62}
\end{equation}%
\ This result implies that, indeed, up to a constant, the obtained
wavefunction should be related to the Rogers-Szego polynomial. The obtained
relation is nontrivial nevertheless. We would like to discuss this
nontriviality in some detail now.

Following [51,53], let $q=e^{-c^{2}},$ where $c$ is some nonegative number.
Introduce the distributed Gaussian polynomials via%
\begin{equation}
\Phi _{n}(x)=\sum\limits_{k=0}^{\infty
}C_{k}^{n}(q)(-1)^{k}q^{-k/2}q^{(x-k)^{2}}.  \tag{63}
\end{equation}%
These polynomials satisfy the following orthogonality relation:%
\begin{equation}
\int\limits_{-\infty }^{\infty }\Phi _{n}(x)\Phi _{m}(x)dx=\left\Vert \Phi
_{n}(x)\right\Vert ^{2}\delta _{mn}  \tag{64}
\end{equation}%
with\footnote{%
Notice that $\alpha =\left( \int\limits_{-\infty }^{\infty
}dxq^{2x^{2}}\right) ^{-\frac{1}{2}}=\left( \frac{\pi }{2c^{2}}\right) ^{-%
\frac{1}{4}}$} 
\begin{equation}
\left\Vert \Phi _{n}(x)\right\Vert =\left( \frac{\pi }{2c^{2}}\right) ^{%
\frac{1}{4}}q^{-\frac{n}{2}}\sqrt{(q,q)_{n}}.  \tag{65}
\end{equation}%
This result calls for a change in normalization of $\Phi _{n}(x),$ i.e$%
.,\phi _{n}(x)=\frac{\Phi _{n}(x)}{\left\Vert \Phi _{n}(x)\right\Vert }.$
Under such conditions $\phi _{n}(x)$ coincides with $\Psi _{n\text{ }}(x),$
provided that $w(x)=1.$ Introduce new variable $u=q^{-2x},$ and consider a
shift: $\Phi _{n}(x)\rightarrow \Phi _{n}(x-s).$ Using (63), we can write 
\begin{equation}
\Phi _{n}(x-s)=u^{s}\exp \{-\left( \ln u\right) ^{2}/(-4\ln q)\}P_{n}(u;s), 
\tag{66a}
\end{equation}%
where 
\begin{equation}
P_{n}(u;s)=\sum\limits_{k=0}^{\infty
}C_{k}^{n}(q)(-1)^{k}q^{-k/2}q^{(s+k)^{2}}u^{k}.  \tag{66b}
\end{equation}%
The orthogonality relation (64) is converted then into 
\begin{equation}
\int\limits_{0}^{\infty }duu^{2s-1}\exp \{-\left( \ln u\right) ^{2}/(-4\ln
q)\}P_{n}(u;s)P_{m}(u;s)=\delta _{mn}.  \tag{67}
\end{equation}%
In view of (67), consider now a special case: $s=1/2$. Then, the weight
function is known as $\QTR{sl}{lognormal}$ distribution and polynomials $%
P_{n}(u;1/2)$ \ (up to a constant ) are known as Stieltjes-Wigert (S-W)
polynomials. Their physical relevance will be discussed below in Subsection
4.6. In the meantime, we introduce the Fourier transform of $f(x)$ in the
usual way as 
\begin{equation}
\int\limits_{-\infty }^{\infty }dx\exp (2\pi i\theta x)f(x)=f(\theta ) 
\tag{68}
\end{equation}%
Then, the Parseval relation implies: 
\begin{equation}
\int\limits_{-\infty }^{\infty }\Phi _{n}(x)\Phi
_{m}(x)dx=\int\limits_{-\infty }^{\infty }\Phi _{n}(\theta )\Phi _{m}(\theta
)dx=\left\Vert \Phi _{n}(x)\right\Vert ^{2}\delta _{mn},  \tag{69a}
\end{equation}%
causing%
\begin{equation}
\Phi _{n}(\theta )=\left( \frac{\pi }{c^{2}}\right) ^{\frac{1}{4}}\exp
(-\left( \pi /c\right) \theta ^{2})\sum\limits_{k=0}^{\infty
}C_{k}^{n}(q)(-q^{-\frac{1}{2}}e^{2\pi i\theta })^{k}.  \tag{69b}
\end{equation}%
Comparison between these results and Eq.(32) produces%
\begin{equation}
\int\limits_{-\infty }^{\infty }H_{n}(-q^{-\frac{1}{2}}e^{-2\pi i\theta
};q)H_{m}(-q^{-\frac{1}{2}}e^{-2\pi i\theta };q)\exp (-2\left( \pi /c\right)
\theta ^{2})=\left( \frac{c}{2\pi }\right) ^{\frac{1}{2}}q^{-n}(q,q)_{n}%
\delta _{mn}  \tag{70a}
\end{equation}%
which can be alternatively rewritten as 
\begin{equation}
\int\limits_{0}^{1}H_{n}(-q^{-\frac{1}{2}}e^{-2\pi i\theta })H_{m}(-q^{-%
\frac{1}{2}}e^{-2\pi i\theta })\vartheta _{3}(2\pi \theta ;q)d\theta
=q^{-n}(q,q)_{n}\delta _{mn}  \tag{70b}
\end{equation}%
with $\vartheta _{3}(\theta ,q)=\sum\limits_{n=-\infty }^{\infty
}q^{n^{2}/2}e^{in\theta }$ . That is $\vartheta _{3\text{ \ }}$is one of the
Jacobi's theta functions. In order to use the obtained results, it is useful
to compare them against those, known in literature already, e.g. see [54]. \
Our Eq.(70a) is in agreement with (5) of [54] if we make identifications: $%
\kappa =\pi $ and $c=\sqrt{2}\kappa ,$ where $\kappa $ is the parameter
introduced in this reference. With help of such an identification we can
proceed with comparison. For this purpose, following [54] we introduce yet
another generating function 
\begin{equation}
S_{n}(t;q):=\sum\limits_{k=0}^{n}\left[ 
\begin{array}{c}
n \\ 
k%
\end{array}%
\right] _{q}q^{k^{2}}t^{k}  \tag{71}
\end{equation}%
so that the S-W polynomials can be written now as 
\begin{equation}
\mathcal{\tilde{S}}_{n}(t;q)=(-1)^{n}q^{\frac{n}{2}}(\sqrt{(q,q)_{n}}%
)^{-1}P_{n}(t;\frac{1}{2})\equiv (-1)^{n}q^{\frac{2n+1}{{}}}(\sqrt{(q,q)_{n}}%
)^{-1}S_{n}(-q^{\frac{1}{2}}t;q),  \tag{72}
\end{equation}%
in accord with Ref.[55], page 197, provided that $0<q<1.$ Comparison between
generating functions (32) and (72) allows us to write as well%
\begin{equation}
S_{n}(t;q^{-1})=H_{n}(tq^{-n};q),\text{ or, equivalently, }%
H_{n}(t;q^{-1})=S_{n}(q^{-n}t;q).  \tag{73}
\end{equation}%
Using this result we can rewrite the recursion relation (33) for $H_{n}(t;q)$
in terms of the recursion relation for $S_{n}(t;q)$ if needed and then to
repeat all the arguments with creation and annihilation operators, etc. For
the sake of space, we leave this option as an exercise for our readers.
Instead, to finish our discussion we would like to show how the obtained
polynomials reduce to the usual Hermite polynomials in the limit $%
q\rightarrow 1^{-}.$ For this purpose we would like to demonstrate that the
recursion relation (33) is actually the recursion relation for the \textsl{%
continuous} $q-$Hermite polynomials [56,57]. This means that we have to
demonstrate that under some conditions (to be specified) the recursion (33)
is equivalent to 
\begin{equation}
2xH_{n}(x\mid q)=H_{n+1}(x\mid q)+(1-q^{n})H_{n-1}(x\mid q).  \tag{74}
\end{equation}%
known for q-Hermite polynomials. To demonstrate the equivalence, we assume
that $x=\cos \theta $ in (74) and then, let $z=e^{i\theta }.$ Furthermore,
we assume that 
\begin{equation}
H_{n}(x\mid q)=z^{n}H_{n}(z^{-2};q)  \tag{75}
\end{equation}%
allowing us to obtain%
\begin{equation}
(z+z^{-1})z^{n}H_{n}=z^{n+1}H_{n+1}+(1-q^{n})z^{n-1}H_{n-1}.  \tag{76}
\end{equation}%
Finally, we set z$^{-1}=\sqrt{t}$ which brings us back to Eq.(33). This
time, however, we can use results known in literature for $q-$Hermite
polynomials [\textbf{41-43}] in order to obtain at once 
\begin{equation}
\lim_{q\rightarrow 1^{-}}\frac{H_{n}(x\sqrt{\frac{1-q}{2}}\mid q)}{\sqrt{%
\frac{1-q}{2}}}=H_{n}(x),  \tag{77}
\end{equation}%
where $H_{n}(x)$ are the standard Hermitian polynomials. In view of
Eq.s(72),(73), not surprisingly, the S-W polynomials are also reducible to $%
H_{n}(x)$. Details can be found in the same references.

\subsection{ASEP, q-deformed harmonic oscillator and spin chains}

In this subsection we would like to connect the results obtained thus far
with that for XXX and XXZ spin chains. Although a connection \ with XXX \
spin chain was established already at the beginning of this section, it is
extremely helpful to arrive at the same conclusions using alternative
(physically inspired) arguments and methods. To understand the logic of our
arguments we encourage our readers to read Appendix\ at this point. In it we
provide a self contained summary of results related to the asymmetric simple
exclusion process (ASEP), especially emphasizing its connection with static
and dynamic properties of XXX\ and XXZ spin chains.

ASEP was discussed in high energy physics literature, e.g. see Ref.[58], in
connection with the random matrix ensembles. To avoid repeats, we would like
to use the results of Appendix \ in order\ to consider a steady-state regime
only. To be in \ accord with literature on ASEP, we complicate matters by \
imposing some nontrivial boundary conditions.

In the steady -state regime Eq.(A.12) of Appendix acquires the form : $%
SC=\Lambda $. Explicitly, 
\begin{equation}
SC=p_{L\text{ }}ED-p_{R}DE.  \tag{78}
\end{equation}%
In the steady-state regime, the operator $S$ becomes an arbitrary c-number
[59]. In view of this, following Sasamoto [60] we rewrite (78) as 
\begin{equation}
p_{R}DE-p_{L}ED=\zeta \left( D+E\right) .  \tag{79}
\end{equation}%
Such operator equation should be supplemented by the boundary conditions
which are chosen to be as 
\begin{equation}
\alpha <W\mid E=\zeta <W\mid \text{ and }\beta D\mid V>=\zeta \mid V>. 
\tag{80}
\end{equation}%
The normalized steady-state probability for some configuration $\mathcal{C}$
can be written \ now as 
\begin{equation}
P(\mathcal{C})=\frac{<W\mid X_{1}X_{2}\cdot \cdot \cdot X_{N}\mid V>}{<W\mid
C^{N}\mid V>}  \tag{81}
\end{equation}%
with \ the operator $X_{i}$ being either $D$ or $E$ depending on wether the $%
i-$th site is occupied or empty. To calculate $P(\mathcal{C})$ we need to
determine $\zeta $ while assuming parameters $\alpha $ and $\beta $ to be
assigned. In Appendix we demonstrate that it is possible to equate $\zeta $
to one \ so that, in agreement with Ref.\textbf{[}61], we obtain the
following representation for $D$ and $E$ operators:%
\begin{equation*}
D=\frac{1}{1-q}+\frac{1}{\sqrt{1-q}}a,\ \text{\ }E=\frac{1}{1-q}+\frac{1}{%
\sqrt{1-q}}a^{+}
\end{equation*}%
converting equation (78) into (52d). In view of this mapping into $q-$%
deformed oscillator algebra, we can expand both vectors $\mid V>$ and $%
<W\mid $ into Fourier series, e.g. $\mid V>=\sum\nolimits_{m}\Omega
_{m}(V)\mid m>$ where, using (57), we put $\mid m>=\Psi _{m}.$ By combining
equations (57) and (80) and results of Appendix we obtain the following
recurrence equation for $\Omega _{n}:$ 
\begin{equation}
\Omega _{n}(V)(\frac{1-q}{\beta }-1)=\Omega _{n+1}(V)\sqrt{1-q^{n+1}}. 
\tag{82}
\end{equation}%
Following [\textbf{47}], we assume that $<0\mid V>=1.$ Then, the above \
recurrence produces 
\begin{equation*}
\Omega _{n}(V)=\frac{v^{n}}{\sqrt{(q,q)_{n}}},
\end{equation*}%
with \ parameter $v=\frac{1-q}{\beta }-1.$ Analogously, we obtain: 
\begin{equation*}
\Omega _{n}(W)=\frac{w^{n}}{\sqrt{(q,q)_{n}}},
\end{equation*}%
with $w=\frac{1-q}{\alpha }-1.$ Obtained results exhibit apparently singular
behavior for $q\rightarrow 1^{-}.$ These singularities are only apparent
since they cancel out when one computes quantities of\ physical interest
discussed in both [62] and [63]. As results of Appendix indicate, such a
crossover is also nontrivial physically since it involves careful treatment
of the transition from XXZ to XXX antiferromagnetic spin chains. Hence, the
results obtained thus far enable us to connect the partition function (27)
(or (32)) with either XXX or XXZ spin chains but are not yet sufficient for
making an unambiguous choice between these two models. This task is
accomplished below, in the rest of this section.

\subsection{Crossover between the XXZ and XXX spin chains: connections with
the KPZ and EW equations and the lattice Liouville model}

\bigskip

Following Derrida and Malick [63], we notice that ASEP is the lattice
version of the famous Kradar-Parisi-Zhang \ (KPZ) equation [64]. The
transition $q\rightarrow 1^{-}$ corresponds to a transition (in the sense of
renormalization group analysis) from the regime of ballistic deposition
whose growth is described by the KPZ \ equation to another regime described
by the Edwards-Wilkinson (EW) \ equation. In the context of ASEP (that is
microscopically) such a transition is discussed in detail in [65].
Alternative treatment is given in [63]. The task of obtaining the KPZ or EW
equations from those describing the ASEP is nontrivial and was accomplished
only very recently [66,67]. It is essential for us that in doing so the
rules of constructing the restricted solid-on -solid (RSOS) models were
used. From the work by Huse [68] it is known that such models can be found
in four thermodynamic regimes. The crossover from the regime III to regime
IV is described by the\ critical exponents of the Friedan, Qui and Shenker
unitary CFT series [69]. The crossover from the regime III to regime IV
happens to be relevant to the crossover from KPZ to EW regime as we would
like to explain now.

As results of \ our Appendix\ indicate, the truly asymmetric \ simple
exclusion process is associated with the XXZ model at the microscopic level
and with the KPZ equation/model at the macroscopic level. Accordingly, the
symmetric exclusion process is associated with the XXX model at the
microscopic level and with the EW equation/model \ at the macroscopic level.
At the level of the Bethe ansatz for open XXZ chain with boundaries full
details of the crossover from the KPZ to EW regime were exhaustively worked
out only very recently [70]. For the purposes of this work it is important
to notice that for certain values of parameters the Hamiltonian of open XXZ
spin chain model \footnote{%
That is Eq.(1.3) of [70].} \ with boundaries can be brought \ to the
following canonical form%
\begin{equation}
H_{XXZ}=\frac{1}{2}[\sum\limits_{j=1}^{N-1}(\sigma _{j}^{x}\sigma
_{j+1}^{x}+\sigma _{j}^{y}\sigma _{j+1}^{y}+\frac{1}{2}(q+q^{-1})\sigma
_{j}^{z}\sigma _{j+1}^{z})+\frac{1}{2}(q-q^{-1})(\sigma _{1}^{z}-\sigma
_{N}^{z})].  \tag{83}
\end{equation}%
In the case of ASEP we put $q=\sqrt{p_{R}/p_{L}}$ so that for physical
reasons parameter $q$ is not complex. However, mathematically, we can allow $%
q$ \ to become complex. In particular, following Pasquer and Saleur [71], we
can let $q=e^{i\gamma }$ with $\gamma =\dfrac{\pi }{\mu +1}.$ For such
values of $q$ use of finite scaling analysis applied to the spectrum of the
above defined Hamiltonian produces \ the central charge 
\begin{equation}
c=1-\frac{6}{\mu (\mu +1)}\text{ , }\mu =2,3,....  \tag{84}
\end{equation}%
of the unitary CFT \ series. Furthermore, if $e_{i}$ is generator of the
Temperley-Lieb algebra\footnote{%
That is $e_{i}^{2}=e_{i},$ $e_{i}e_{i+1}e_{i}=q^{-1}e_{i}$ and $%
e_{i}e_{j}=e_{j}e_{i}$ for $\left\vert i-j\right\vert \geq 2.$}, then $%
H_{XXZ}$ can be rewritten as [72] 
\begin{equation}
H_{XXZ}=-\sum\limits_{j=1}^{N-1}[e_{i}-\frac{1}{4}(q+q^{-1})].  \tag{85}
\end{equation}%
This fact allows us to make immediate connections with quantum groups and
theory of knots and links. Below, in Section 5 we shall discuss briefly
different arguments leading to analogous conclusions.

The results just described allow us to connect CFT with the\ exactly
integrable \ lattice models. In view of this, one can pose the following
question: Given the connection we just described, can we write down
explicitly the corresponding path integral string-theoretic models
reproducing results of exactly integrable lattice models at and away from
criticality? Before providing \ the answer in the following subsection, we
would like to conclude this subsection with a partial answer. In particular,
we would like to mention the work by Faddeev and Tirkkonen [73] connecting
the \textsl{lattice} Liouville model with the spin 1/2 XXZ chain. Based on
this result, it should be clear that in the region \ $c\leq 1$ it is indeed
possible by using the combinatorial analysis described above to make a link
between the continuum and discrete Liouville theories\footnote{%
The matrix $c=1$ theories will be discussed separately below.}. It can be
made in such a way that, at least at crtiticality, the results of exactly
integrable 2 dimensional models are in agreement with those which are
obtainable field- theoretically. The domain $c>1$ is physically meaningless
because the models (\textsl{other} \textsl{than string-theoretic}) we
discussed in this section loose their physical meaning in this region. This
conclusion \ will be further reinforced in the next subsection.

\subsection{ASEP, vicious random walkers and string models}

We have discussed at length the role of vicious random walkers in our
treatment of the Kontsevich-Witten (K-W) model [74]. Forrester [75] noticed
that the random turns vicious walkers model is just a special case of ASEP.
Further details on connections between the ASEP, vicious walkers, KPZ and
random matrix theory can be found in the paper by Sasamoto [76]. In the
paper by Mukhi [77] it is emphasized that while the K-W model is the matrix
model representing $c<1$ bosonic string, \ the Penner matrix model with
imaginary coupling constant is representing $c=1$ bosonic\ Euclidean string
on the cylinder of \ (self-dual) radius $R=1\footnote{%
This was initially demonstrated by Distler and Vafa [78].}.$ Furthermore,
Ghoshal and Vafa [79] have demonstrated that $c=1,R=1$ string is dual to the
topological string on a conifold singularity. We shall \ briefly discuss
this connection below. Before doing so, \ it is instructive to discuss the
crossover from $c=1$ to $c<1$ string models in terms of vicious walkers. To
do so we shall use some results from our work on K-W model and from the
paper by Forrester [75].

Thus, we would like to consider a planar lattice where at the beginning we
place only one directed path $P$: from $(a,1)$ to $(b,N)$\footnote{%
Very much in the same way as \ we have discussed in Section 2.}. The
information about this path can be encoded into multiset $Hor_{y}(P)$ of $y$%
-coordinates of the horizontal steps of $P$. Let now%
\begin{equation}
w(P)=\prod\limits_{i=Hor_{y}(P)}x_{i}.  \tag{86}
\end{equation}%
Using these definitions, the extension of these results to an assembly of
directed random vicious walkers is given as a product: $W(\hat{P})\equiv
w(P_{1})\cdot \cdot \cdot w(P_{k}).$ Finally, the generating function for an
assembly of such walkers is given by%
\begin{equation}
h_{b-a}(x_{1},...,x_{N})=\sum\limits_{\hat{P}}W(\hat{P}),  \tag{87}
\end{equation}%
where $W(\hat{P})$ is made of monomials of the type $%
x_{1}^{m_{1}}x_{2}^{m_{2}}\cdot \cdot \cdot x_{N}^{m_{N}}$ provided that $%
m_{1}+\cdot \cdot \cdot +m_{N}=b-a.$ The following theorem discussed in our
work, Ref[\textbf{60}], is of central importance for calculation of such
defined generating function.

Given integers $0<a_{1}<\cdot \cdot \cdot <a_{k\text{ \ }}$and $%
0<b_{1}<\cdot \cdot \cdot <b_{k}$ , let \textbf{M}$_{i,j}$ be the $k\times k$
matrix \textbf{M}$_{i,j}=h_{b_{j}-a_{i}}(x_{1},...,x_{N})$ then, 
\begin{equation}
\det \mathbf{M}=\sum\limits_{\hat{P}}W(\hat{P}),  \tag{88}
\end{equation}%
where the sum is taken over all sequences ($P_{1},...,P_{k})\equiv \hat{P}$
of nonintersecting lattice paths $P_{i}:(a_{i},1)\rightarrow
(b_{i},N),i=1-k. $

Let now $a_{i}=i$ and $b_{j}=\lambda _{i}+j$ so that $1\leq i,j\leq k$ with $%
\lambda $ being a partition of $N$ with $k$ parts then, $\det \mathbf{M}%
=s_{\lambda }(x_{1},...,x_{N}),$ where $s_{\lambda }(\mathbf{x})$ is the
Schur polynomial. In our work [74] we demonstrated that in the limit $%
N\rightarrow \infty $ such defined Schur polynomial coincides with the
partition (generating) function for the Kontsevich model. \ To obtain
results of Forrester requires\ us to apply some additional efforts. These
are worth discussing.

Unlike the K-W case, this time, we need to discuss \ the \textsl{continuous}
random walks in the plane. Let $x$-coordinate represent "space" while $y$%
-coordinate- "time". If initially ($t=0$) we had $k$-walkers in positions $%
-L $ $<x_{1}<x_{2}<\cdot \cdot \cdot <x_{k\text{ }}<L,$ the same order
should persist $\forall $ $t>0$. At each tick of the clock each walker is
moving either to the right or to the left with equal probability $p$ (that
is we are in the regime appropriate for the XXX spin chain in the ASEP
terminology). As before, let \textbf{x}$_{0}=(x_{1,0},...,x_{k,0})$ be the
initial configuration of $k-$walkers and \textbf{x}$%
_{f}=(x_{1,f},...,x_{k,f})$ be the final configuration at time $t$. To
calculate the total number of walks starting at time $t=0$ at \textbf{x}$%
_{0} $ and ending at time $t$ at \textbf{x}$_{f}$ we need to know the
probability distribution $W_{k}(\mathbf{x}_{0}\rightarrow \mathbf{x}_{f};t)$
that the walkers proceed without bumping into each other. Should these
random walks be totally uncorrelated, we would obtain for the probability
distribution the standard Gaussian result%
\begin{equation}
W_{k}^{0}(\mathbf{x}_{0}\rightarrow \mathbf{x}_{f};t)=\frac{\exp \{-(\mathbf{%
x}_{f}-\mathbf{x}_{0})^{2}/2Dt\}}{\left( 2\pi Dt\right) ^{k/2}}.  \tag{89}
\end{equation}%
In the present case the walks are restricted (correlated) so that the
probability should be modified. This modification can be found in the work
by Fisher and Huse [80]. These authors obtain%
\begin{equation}
W_{k}(\mathbf{x}_{0}\rightarrow \mathbf{x}_{f};t)=U_{k}(\mathbf{x}_{0},%
\mathbf{x}_{f};t)\frac{\exp \{-(\mathbf{x}_{f}^{2}+\mathbf{x}_{0}^{2})/2Dt\}%
}{\left( 2\pi Dt\right) ^{k/2}}  \tag{90}
\end{equation}%
with%
\begin{equation}
U_{k}(\mathbf{x}_{0},\mathbf{x}_{f};t)=\sum\limits_{g\in S_{k}}\varepsilon
(g)\exp [\frac{(\mathbf{x}_{f}\cdot g\mathbf{x}_{0})}{Dt}].  \tag{91}
\end{equation}%
In this expression $\varepsilon (g)=\pm 1,$ and the index $g$ runs over all
members of the symmetric group $S_{k}$. Mathematically, following Gaudin
[81], this problem can be looked upon as a problem of a random walk inside
the $k-$dimensional kaleidoscope (Weyl cone) usually complicated by
imposition of some boundary conditions on walks at the walls of the cone.
Connection of such random walk problem with random matrices was discussed by
Grabiner [82] whose results were \ very recently improved and generalized by
Krattenthaller [83]. In the work by de Haro some applications of Grabiner's
results to high energy physics were considered [84]. Here we would like to
approach the same class of problems based on the results obtained in this
paper. In particular, some calculations made in [80] indicate that for $%
L\rightarrow \infty $ with accuracy up to $O$ ($L^{2}/Dt)$ it is possible to
rewrite $U_{k}(\mathbf{x}_{0},\mathbf{x}_{f};t)$ as follows:%
\begin{equation}
U_{k}(\mathbf{x}_{0},\mathbf{x}_{f};t)\simeq const\Delta (\mathbf{x}%
_{f})\Delta (\mathbf{x}_{0})/\left( Dt\right) ^{n_{k}}+O(L^{2}/Dt)  \tag{92}
\end{equation}%
with $n_{k}=\left( 1/2\right) k(k-1)$ and $const=1/1!2!\cdot \cdot \cdot
(k-1)!$ and $\Delta (\mathbf{x})$ being the Vandermonde determinant, i.e.%
\begin{equation}
\Delta (\mathbf{x})=\prod\limits_{i<j}(x_{i}-x_{j}).  \tag{93}
\end{equation}%
Next, from standard texts in probability theory it is known that \textsl{%
non-normalized} expression for $W_{k}^{0}(\mathbf{x}_{0}\rightarrow \mathbf{x%
}_{f};t)$ in the limit of long times provides the number of random walks of $%
n$ steps (since $n\rightleftarrows t)$ from point $\mathbf{x}_{0}$ to point 
\textbf{x}$_{f}$. Hence, the same must be true for $W_{k}(\mathbf{x}%
_{0}\rightarrow \mathbf{x}_{f};t)$ and, therefore, $W_{k}(\mathbf{x}%
_{0}\rightarrow \mathbf{x}_{f};t)\approx \det \mathbf{M.}$ Consider such
walks for which $\mathbf{x}(t=0)\equiv \mathbf{x}_{0}=\mathbf{x}%
(t=t_{f})\equiv \mathbf{x}_{f}.$ Then, using (90) and (92) we obtain the
probability distribution for the Gaussian unitary ensemble [85], i.e.%
\begin{equation}
W_{k}(\mathbf{x}_{0}=\mathbf{x}_{f};t)=const\text{ }\Delta ^{2}(\mathbf{x}%
)\exp (-\mathbf{x}^{2}).  \tag{94}
\end{equation}%
Some additional manipulations (described in our work [74]) using this
ensemble lead directly to the K-W matrix model. Forrester [75] considered a
related quantity- the probability that \textsl{all} $k$ vicious walkers will
survive at time $t_{f}.$ To obtain this probability requires integration of $%
W_{k}(\mathbf{x}_{0}\rightarrow \mathbf{x}_{f};t)$ over the simplex $\Delta $
defined by $-\ L<x_{1}<x_{2}<\cdot \cdot \cdot <x_{k\text{ }}<L\footnote{%
Such type of integration is described in detail in our papers, Parts I and
II, from which it follows that in the limit $L\rightarrow \infty $ such a
simplex integration can be replaced by the usual integration, i.e $%
\int\nolimits_{\Delta }\prod\limits_{i}^{k}dx_{i}\cdot \cdot \cdot
\rightleftarrows \frac{1}{k!}\int\limits_{-\infty }^{\infty
}\prod\limits_{i}^{k}dx_{i}\cdot \cdot \cdot $ in accord with Forrester.}.$
Without loss of generality, it is permissible to use $W_{k}(\mathbf{x}_{0}=%
\mathbf{x}_{f};t)$ instead of $W_{k}(\mathbf{x}_{0}\rightarrow \mathbf{x}%
_{f};t)$ in calculating such a probability. Then, the obtained result
coincides (up to a constant) with the partition function for topological
gravity $\mathcal{Z}(g),$ Eq.(3.1) of [86,87]\footnote{%
Since the hermitian matrix model given by Eq.(3.1) is just a partition
function for the Gaussian unitary ensemble [85].}). Furthermore, such
defined partition function can be employed to reproduce back the Hermite
polynomial $H_{k}(x)$ defined by (77) which has an interpretation as the
wavefunction(amplitude) of the FZZT $D-$brane [86,87]. Specifically, we have%
\begin{eqnarray}
&<&\det (x-M)>=\left( \frac{g}{4}\right) ^{\frac{n}{2}}H_{k}(x\sqrt{\frac{1}{%
g}})  \notag \\
&=&\frac{1}{\mathcal{Z}(g)}\int dM\det (x-M)e^{-\frac{1}{g}trM^{2}}. 
\TCItag{95}
\end{eqnarray}%
This expression is a special case of Heine's formula representing monic
orthogonal polynomials through random matrices. In the above formula $k$ is
related to the size of Hermitian matrix $M$ and $g$ is the coupling constant.

Following Forrester [75], the result, Eq.(90), can be treated more
accurately (albeit a bit speculatively) if, in addition to the parameter $D$
we introduce \ another parameter $a$ - the spacing between random walkers at
time $t=0$. Furthermore, if the time direction is treated as space direction
(as it is commonly done for 1d quantum systems in connection with 2d
classical systems), then yet another parameter $\tau (k,t)$ should be
introduced which effectively renormalizes $D$. This eventually causes us to
replace $\mathcal{Z}(g)$ by the following integral (up to a constant)%
\begin{equation}
\mathcal{\hat{Z}}(g)=\prod\limits_{i=1}^{k}\int\limits_{-\infty }^{\infty
}dx_{i}\exp (-\frac{1}{2g}\ln
^{2}x_{i})\prod\limits_{i<j}(x_{i}-x_{j})^{2}\equiv \int dMe^{-\frac{1}{2g}%
tr(\ln M)^{2}}.  \tag{96}
\end{equation}%
Tierz demonstrated [88] that $\mathcal{\hat{Z}}(g)$ (up to a constant) is
partition function of the Chern-Simons (C-S) field theory with gauge group $%
U(k)$ living on 3-sphere $S^{3}.$ Okuyama [87] used (95) in order to get
analogous result for a D-brane amplitude in the C-S model. Using Heine's
formula, he obtained the Stieltjes-Wiegert \ (S-W) polynomial, our Eq.(71),
which can be expressed via Rogers-Szego polynomial according to (73) and,
hence, via the $q-$Hermite polynomial in view of the relation (75). Since in
the limit $q\rightarrow 1^{-}$ the $q-$Hermite polynomial is reducible to
the usual Hermite polynomial according to (77), there should be analogous
procedure in going from the partition function $\mathcal{\hat{Z}}(g)$ to $%
\mathcal{Z}(g).$ Such a procedure can be developed, in principle, by
reversing arguments of Forrester. However, these arguments are much less
rigorous and physically transparent than those used in previous subsection
where we discussed the crossover from XXZ to XXX model. \ In view of the
results presented below, in the following section, we leave the problem of
crossover between the matrix ensembles outside the scope of this paper. To
avoid duplications, we refer our readers to the paper by Okuyama [87] where
details are provided relating our results to the topological A and B -branes.

\section{ Discussion. Gaudin model as linkage between the WZNW model and K-Z
equations}

\ 

We would like to remind to our readers that all results obtained thus far
can be traced back to our Eq.(32) defining the Rogers-Szego polynomial which
\ physically was interpreted as partition function for the ferromagnetic P-F
spin chain\footnote{%
The antiferromagnetic version of P-F spin chain is easily obtainable from
this ferromagnetic version as discussed in Section 3.}. In previous section
numerous attempts\ were made to connect this partition function to various
known string models, even though already in Section 4.1 we came to the
conclusion that in the limit of infinitely long chains the antiferromagnetic
version of P-F spin chain can be \ replaced by the spin 1/2
antiferromagnetic XXX chain. If this is so, then from literature it is known
that the behavior of such spin chain is described by the $SU_{1}(2)$ WZNW
model [44]. Hence, at the physical level of rigor\ the problem of connecting
Veneziano amplitudes to physical models can be considered as completely
solved. Here we argue that at the mathematical level of rigor this is not
quite so yet. This conclusion concerns not only problems discussed in this
paper but, in general, the connection between WZNW models, spin chains and
K-Z equations.

It is true that the K-Z equations and WZNW model \ are inseparable from each
other [44] but the extent\ to which spin chains can be directly linked to
both \ the WZNW models and K-Z equations remains to be explained. Following
Varchenko [89], we notice that the link between the K-Z equations and WZNW
models can be made\ only with help of\ the Gaudin model, while the
connection with spin chains can be made only by using the \textsl{quantum}
version of the K-Z equation. Such quantized version of \ the K-Z equation 
\textsl{is not} immediately connected with the standard WZNW model \ as
discussed in many places [89\textbf{, }90]. For the sake of space we refer
our readers to these references and to our latest work connecting Veneziano
amplitudes with Gaudin models [27] in which we also discuss in detail
physical implications of such a connection.

\bigskip \medskip

\bigskip \textbf{Appendix. } \textbf{Basics of ASEP}$\ \ $

\ 

\textbf{A.1.} \textbf{Equations of motion and spin chains}

\ 

The one dimensional asymmetric simple exclusion process (ASEP) had been
studied for some time [91]. The purpose of this Appendix is to summarize the
key features of this process which are of immediate relevance to the content
of this paper. To this purpose, following Sch\"{u}tz [92], we shall briefly
describe the ASEP with sequential updating. Let $B_{N}:=\{x_{1},...,x_{N}\}$
be the set of sites of one dimensional lattice arranged at time $t$ in such
a way that $x_{1}<x_{2}$\ $<\cdot \cdot \cdot <x_{N}$. It is expected that
each time update will not destroy this order.

\ Consider first the simplest case of $N=1$. Let $p_{R}$ $(p_{L})$ be the
probability of a particle located at the site $x$ to move to the right
(left) then, after transition to continuous time, the master equation for
the probability $P(x;t)$ can be written as follows%
\begin{equation}
\frac{\partial }{\partial t}P(x;t)=p_{R}P(x-1;t)+p_{L}P(x+1;t)-P(x;t). 
\tag{A.1}
\end{equation}%
Assuming that $P(x;t)=\exp (-\varepsilon t)P(x)$ so that that

$P(x;t)=\int\limits_{0}^{2\pi }dp\exp (-\varepsilon t)f(p)\exp (ipx),$ $p\in
\lbrack 0,2\pi ),$ we obtain the dispersion relation for the energy $%
\varepsilon (p):$%
\begin{equation}
\varepsilon (p)=p_{R}(1-e^{-ip})+p_{L}(1-e^{ip}).  \tag{A.2}
\end{equation}%
The initial condition $P(x;0)=\delta _{x,y}$ determines the amplitude $%
f(p)=e^{-ipy}/2\pi $ and finally yields%
\begin{equation}
P(x;t\shortmid y;0)=\frac{1}{2\pi }\int\limits_{0}^{2\pi }dpe^{-\varepsilon
(p)t}e^{-ipy}e^{ipx}=e^{-(q+q^{-1})Dt}q^{x-y}I_{x-y}(2Dt),  \tag{A.3}
\end{equation}%
where $q=\sqrt{p_{R}/p_{L}}$, $D=\sqrt{p_{R}p_{L}}$ and $I_{n}(2Dt)$ is the
modified Bessel function. These results can be easily extended to the case $%
N=2$. Indeed, for this case \ we obtain the following equation of motion%
\begin{eqnarray}
\varepsilon P(x_{1},x_{2})
&=&-p_{R}(P(x_{1}-1,x_{2})+P(x_{1},x_{2}-1)-2P(x_{1},x_{2}))  \notag \\
&&-p_{L}(P(x_{1}+1,x_{2})+P(x_{1},x_{2}+1)-2P(x_{1},x_{2}))  \TCItag{A.4}
\end{eqnarray}%
which should be supplemented by the boundary condition%
\begin{equation}
P(x,x+1)=p_{R}P(x,x)+p_{L}P(x+1,x+1)\text{ }\forall x.  \tag{A.5}
\end{equation}%
Imposition of this boundary condition allows us to look for a solution of
(A.4) in the\ (Bethe ansatz) form%
\begin{equation}
P(x_{1},x_{2})=A_{12}e^{ip_{1}x_{1}}e^{ip_{2}x_{2}}+A_{21}e^{ip_{2}x_{1}}e^{ip_{1}x_{2}}
\tag{A.6}
\end{equation}%
yielding $\varepsilon (p_{1},p_{2})=\varepsilon (p_{1})+\varepsilon (p_{2}).$
Use of the boundary condition (A.5) fixes the ratio (the S-matrix) $%
S_{12}=A_{12}/A_{21}$ as follows:%
\begin{equation}
S(p_{1},p_{2})=-\frac{p_{R}+p_{L}e^{ip_{1}+ip_{2}}-e^{ip_{1}}}{%
p_{R}+p_{L}e^{ip_{1}+ip_{2}}-e^{ip_{2}}}.  \tag{A.7}
\end{equation}%
To connect this result with the quantum spin chains,\ we consider the case
of symmetric hopping first. In this case we have $p_{R}=p_{L}=1/2$ so that
(A.7) is reduced to 
\begin{equation}
S_{XXX}(p_{1},p_{2})=-\frac{1+e^{ip_{1}+ip_{2}}-2e^{ip_{1}}}{%
1+e^{ip_{1}+ip_{2}}-2e^{ip_{2}}}  \tag{A.8}
\end{equation}%
from this result we can recognize the $S$ matrix for the XXX spin 1/2
Heisenberg ferromagnet [81]. If $p_{R}\neq p_{L}$, to bring (A.7) in
correspondence with the spin chain $S-$ matrix requires additional efforts.
Following Gwa and Spohn [93] we replace the complex numbers $e^{ip_{1}}$ and 
$e^{ip_{2}}$ in (A.7) respectively by z$_{1}$ and z$_{2}.$ In such a form \
(A.7) exactly \ coincides with the S-matrix \ obtained by Gwa and Spohn%
\footnote{%
E.g. see their Eq.(3.5).}. After this we can rescale z$_{i}$ $(i=1,2)$ as
follows: $z_{i}=\sqrt{\frac{q}{p}}\tilde{z}_{i}.$ Substitution of such an
ansatz into (A.7) leads to the result 
\begin{equation}
S_{XXZ}(\tilde{z}_{1},\tilde{z}_{2})=-\frac{1+\tilde{z}_{1}\tilde{z}%
_{2}-2\Delta \tilde{z}_{1}}{1+\tilde{z}_{1}\tilde{z}_{2}-2\Delta \tilde{z}%
_{2}},  \tag{A.9}
\end{equation}%
provided that $2\Delta =1/\sqrt{p_{L}p_{R}}$. For $p_{R}=p_{L}=1/2$ we
obtain $\Delta =1$ as required for the XXX chain\footnote{%
It should be noticed though that such a parametrization is not unique. For
instance, following [70] it is possible to choose a slightly different
parametrization, e.g. $\Delta =-\frac{1}{2}(q+q^{-1}),$ where $q=\sqrt{%
p_{R}/p_{L}}.$} If, however, $p_{R}\neq p_{L},$ then, the obtained $S$%
-matrix \ coincides with that known for the XXZ model [81] if we again
relabel \~{z}$_{i}$ by $e^{ip_{i}}$ which is always permissible since the
parameter $p$ is determined by the Bethe equations (to be discussed below)
anyway.

In the case of XXZ spin chain it is customary to think about the massless $%
-1\leq \Delta \leq 1$ and massive $\left\vert \Delta \right\vert >1$ regime.
The massless regime describes various CFT discussed in the text while the
massive regime describes massive excitations away from criticality. As
Gaudin had demonstrated [81], for XXZ chain it is sufficient to consider
only $\Delta >0$ domain \ which makes XXZ model perfect for uses in ASEP.
The cases $\Delta =0$ and $\Delta \rightarrow \infty $ also physically
interesting: the first corresponds to the XY model and the second to the
Ising model.

Once the S-matrix is found, \ the $N-$ particle solution can be easily
constructed [92]. For instance, for $N=3$ we write%
\begin{eqnarray*}
\Psi (x_{1},x_{2},x_{3}) &=&\exp
(ip_{1}x_{1}+ip_{2}x_{2}+ip_{3}x_{3})+S_{21}\exp
(ip_{2}x_{1}+ip_{1}x_{2}+ip_{3}x_{3}) \\
&&+S_{32}S_{31}\exp
(ip_{2}x_{1}+ip_{3}x_{2}+ip_{1}x_{3})+S_{21}S_{31}S_{32}\exp
(ip_{3}x_{1}+ip_{2}x_{2}+ip_{3}x_{3}) \\
&&+S_{31}S_{32}\exp (ip_{3}x_{1}+ip_{1}x_{2}+ip_{2}x_{3})+S_{32}\exp
(ip_{1}x_{1}+ip_{3}x_{2}+ip_{2}x_{3}),
\end{eqnarray*}%
etc. This result is used instead of $f(p)$ in (A.3) so that the full
solution is given by 
\begin{equation}
P(x_{1},...,x_{N};t\shortmid y_{1},...,y_{N};0)=\prod\limits_{l=1}^{N}\frac{1%
}{2\pi }\int\limits_{0}^{2\pi }dp_{l}e^{-\varepsilon
(p_{l})t}e^{-ip_{l}y_{l}}\Psi (x_{1},...,x_{N}).  \tag{A.11}
\end{equation}%
The above picture can be refined as follows. First, the particle sitting at $%
x_{i}$ will move to the right(left) only if the nearby site is not occupied.
Hence, the probabilities $p_{R}$ and $p_{L}$ can have values ranging from 0
to 1. For instance, for the totally asymmetric exclusion process (TASEP)
particle can move to the right with probability 1 if the neighboring site to
its right is empty. Otherwise the move is rejected. \ Since under such
circumstances particle can never move to the left, there must be a particle
source located next to the leftmost particle position and the particle sink
located immediately after the rightmost position. After imposition of
emission and absorption rates for these sources and sinks, we end up with
the Bethe ansatz complicated by the imposed boundary conditions. Although in
the case of solid state physics \ these conditions are normally assumed to
be periodic, in the present case, they should be chosen among the solutions
of the Sklyanin \ boundary equations [59\textbf{, }72]. At more intuitive
level of presentation compatible with results just discussed, the Bethe
ansatz for XXZ chain accounting for the boundary effects is given in the
pedagogically written paper by Alcaraz et al [94].

\bigskip\ 

\textbf{A.2.} \textbf{Dynamics of ASEP and operator algebra }

\ 

To make these results useful for the main text, several additional steps are
needed. For this purpose we shall follow works by Sasamoto and Wadati [95]
and Stinchcombe and Sch\"{u}tz [59]. In doing so we rederive many of their
results differently.

We begin with observation that the state of one dimensional lattice
containing $N$ sites can be described in terms of a string of operators $D$
and $E,$ \textbf{\ }where $D$ stands for the occupied and $E$ for empty $k-$%
th position along $1d$ lattice. The non normalized probability (of the type
given in (A.11)) can then be presented as a sum of terms like this $\
<EEDEDDD\cdot \cdot \cdot E>$ \ to be discussed in more detail below.

Let $C=D+E$ be the time- independent\textbf{\ }operator. Then, for the
operator $D$ to be time-dependent the following commutation relations should
hold%
\begin{equation}
SC+\dot{D}C=\Lambda ,  \tag{A.12a}
\end{equation}%
\begin{equation}
CS-C\dot{D}=\Lambda ,  \tag{A.12b}
\end{equation}%
\begin{equation}
\dot{D}D+D\dot{D}=[D,S].  \tag{A.12c}
\end{equation}%
If $\Lambda =p_{L}CD$\textbf{\ -}$p_{R}DC$\textbf{\ +}$\mathbf{(}%
p_{R}-p_{L})D^{2}$ \ (or $\Lambda =p_{L\text{ }}ED-p_{R}DE$ ),\ it is
possible to determine $S$ using equations (A.12) so that we obtain,%
\begin{equation}
\dot{D}=\frac{1}{2}[\Lambda ,C^{-1}],  \tag{A.13a}
\end{equation}%
\textbf{\ }%
\begin{equation}
S=\frac{1}{2}\{\Lambda ,C^{-1}\},  \tag{A.13b}
\end{equation}%
provided that 
\begin{equation}
\Lambda C^{-1}D=DC^{-1}\Lambda  \tag{A.13c}
\end{equation}%
with \{ , \} being an anticommutator.

As before, we want to consider the case $p_{R}=p_{L}$\textbf{\ }$\mathbf{=}%
\frac{1}{2}$ first$.$ This condition leads to $\Lambda =\frac{1}{2}[C,D]$.
It is convenient at this stage to introduce an operator $D_{n}$\textbf{\ =}$%
C^{n-1}$\textbf{\ }$DC^{-n}$\textbf{\ \ }and its Fourier\textbf{\ }transform 
$\mathcal{D}_{p}$\textbf{\ =}$\sum\nolimits_{n}D_{n}\exp (ipn).$ Using
(A.13a) with $\Lambda $ just defined leads to the following equation of
motion for $\mathit{D}_{n}:$%
\begin{equation}
\dot{D}_{n}=\frac{1}{2}[D_{n+1}+D_{n-1}-2D_{n}]  \tag{A.14}
\end{equation}%
to be compared with (A.1). Such a comparison produces at once $\mathcal{D}%
_{p}(t)=\exp (-\varepsilon (p))\mathcal{D}_{p}(0)$\textbf{\ }so that \textbf{%
\ }$\varepsilon (p)=1-\cos p$ as before\footnote{%
E.g. see (A.2) with $p_{L}=p_{R}=1/2$}. Consider now Eq.(A.13c). Under
conditions $p_{R}=p_{L}$\textbf{\ }$\mathbf{=}\frac{1}{2}$ it can be written
as

$CDC^{-1}D\mathbf{\ +}DC^{-1}DC\mathbf{\ =}2D^{2}$\textbf{\ }or,\textbf{\ }%
equivalently, as\footnote{%
Since using definition of $D_{n}$ we have: $D_{n+1}=CD_{n}C^{-1}$ and $%
D_{n-1}=C^{-1}D_{n}C.$}\textbf{\ }%
\begin{equation}
D_{n+1}D_{n}+D_{n}D_{n-1}=2D_{n}D_{n}.  \tag{A.15}
\end{equation}%
Following Gaudin [81], we consider a formal expansion $D_{n}D_{m}$\ \ =$%
\alpha \exp (ip_{1}n+ip_{2}m)+\beta $\ $\exp (ip_{2}n+ip_{1}m)$\ \ and use
it in the previous equation in order to obtain:%
\begin{eqnarray}
&&\alpha \exp (ip_{1}(n+1)+ip_{2}n)+\beta \ \exp (ip_{2}(n+1)+ip_{1}n) 
\notag \\
&&+\alpha \exp (ip_{1}n+ip_{2}(n-1))+\beta \ \exp (ip_{2}n+ip_{1}(n-1)) 
\notag \\
&=&2\alpha \exp (ip_{1}n+ip_{2}n)+2\beta \ \exp (ip_{2}n+ip_{1}n). 
\TCItag{A.16}
\end{eqnarray}%
From here we \ also obtain:\ \ \ 
\begin{equation*}
(\alpha \exp (ip_{1}n+ip_{2}n)(\exp (ip_{1})+\exp (-ip_{2})-2)+\beta \exp
(ip_{1}n+ip_{2}n)(\exp (ip_{2})+\exp (-ip_{1})-2)=0
\end{equation*}%
and, therefore, 
\begin{equation}
S(p_{1},p_{2})\equiv \frac{\alpha }{\beta }=-\frac{1+\exp
(i(p_{1}+p_{2}))-2\exp (ip_{1})}{1+\exp (i(p_{1}+p_{2}))-2\exp (ip_{2})}\exp
(i(p_{2}-p_{1}))  \tag{A.17}
\end{equation}%
to be compared with (A.8).\ An extra factor $\exp (i(p_{2}-p_{1}))$ can be
actually dropped from the $S-$matrix in view of the following chain of
arguments.

\ Introduce the correlation function as follows%
\begin{eqnarray}
P(x_{1},...,x_{N};t &\shortmid &y_{1},...,y_{N};0)\equiv Z_{N}^{-1}Tr[%
\mathit{D}_{1}(t)\cdot \cdot \cdot \mathit{D}_{N}(t)C^{N}]  \notag \\
&=&\prod\limits_{l=1}^{N}\frac{1}{2\pi }\int\limits_{0}^{2\pi
}dp_{l}e^{-\varepsilon (p_{l})t}e^{-ip_{l}y_{l}}\Psi (p_{1},...,p_{N}), 
\TCItag{A.18}
\end{eqnarray}%
\ where $\Psi (p_{1},...,p_{N})=$\ $Z_{N}^{-1}Tr[\mathcal{D}_{p_{1}}(0)\cdot
\cdot \cdot \mathcal{D}_{p_{N}}(t)C^{N}]$ and $Z_{N}=tr[C^{N}].$ \ In
arriving at this result the definition of $\mathcal{D}_{p}(t),$ was used
along with\ the fact that $\mathcal{CD}_{p}\mathcal{C}^{-1}=\ e^{-ip}$\ $%
\mathcal{D}_{p}$. Also, the invariance of the trace under cyclic
permutations and the translational invariance of the correlation function
implying that $\Psi (p_{1},...,$ $p_{N})\neq 0$ only if $\sum\nolimits_{i}$ $%
p_{i}$\ \ $=0$ \ was taken into account. These conditions are sufficient for
obtaining the \ Bethe ansatz equations \ 
\begin{equation}
\exp (ip_{i}N)=\prod\limits_{j=1}^{N}\tilde{S}(p_{i},p_{j})\text{ }\forall
i\neq j,  \tag{A.19}
\end{equation}%
\ where $\tilde{S}(p_{i},p_{j})$ is the same $S-$matrix as in (A.17), except
of the missing factor $\exp (i(p_{i}-p_{j}))$\ which is dropped in view of
translational invariance\footnote{%
Surely, in the case when the effects of boundaries should be accounted, this
factor should be treated depending on the kind of boundary conditions
imposed.}.\ \ 

Extension of these results to the case $p_{R}\neq p_{L}$ is nontrivial.
Because of this, we would like to provide some details not shown in the
cited references. In particular, contrary to claims made in [95], we would
like to demonstrate that the system of equations (A.12) obtained in [59] is
equivalent to the system of equations 
\begin{equation}
\lbrack C,S]=0,  \tag{A.20a}
\end{equation}%
\begin{equation}
C\dot{D}+CT-SD=-p_{L}CD+p_{R}DC,  \tag{A.20b}
\end{equation}%
\begin{equation}
\dot{D}C+DS-TC=p_{L}CD-p_{R}DC,  \tag{A.20c}
\end{equation}%
\begin{equation}
\dot{D}D+D\dot{D}=[T,D]  \tag{A.20d}
\end{equation}%
obtained in [95] with the purpose of describing asymmetric processes.

To make a comparison between (A.12) and (A.20), we notice that (A.20) has
operators $S$ and $T$ \ which cannot be trivially identified with those
present in (A.12). Hence, the task lies in making such an identification.
For this purpose, if we assume that $S$ in (A.20) is the same as in (A.12)
then, \ in view of (A.20a), by subtracting (A.12b) from (A.12a) we obtain: 
\begin{equation}
\dot{D}C+C\dot{D}=0.  \tag{A.21}
\end{equation}%
This leads to either $\dot{D}=C^{-1}\dot{D}C$ or $\dot{D}=-C\dot{D}C^{-1}.$
Therefore, taking into account that, by construction, $C$ is
time-independent, we obtain: $D=$ $-CDC^{-1}+\Theta ,$ where $\Theta $ is
some diagonal time-independent matrix operator.

Next, using these results we multiply (A.20b) from the right by $C^{-1}$ and
(A.20c) by $C^{-1}$ from the left, and add them together in order to arrive
at equation (19) of [95], i.e.

\bigskip 
\begin{equation}
2\dot{D}=p_{R}C^{-1}DC+p_{L}CDC^{-1}-(p_{R}+p_{L})D.  \tag{A.22}
\end{equation}%
Also, by multiplying this result from the right by $D$ we obtain equation
(20) of [95], that is 
\begin{equation}
0=p_{R}DC^{-1}DC+p_{L}CDC^{-1}D-(p_{R}+p_{L})D^{2},  \tag{A.23}
\end{equation}%
provided that $[T,D]=0.$ \ That this is indeed the case can be seen from the
same reference where the following result for $T$ is obtained: 
\begin{equation}
2T=(2+p_{R}-p_{L})D+p_{R}C^{-1}DC-p_{L}CDC^{-1}.  \tag{A.24}
\end{equation}%
Using it, we obtain: $[T,D]=0,$ in view of the fact that $[C^{-1}DC,D]=0$
and $[CDC^{-1},D]=0$ since $D=-CDC^{-1}+\Theta $ as we have\ already
demonstrated. Furthermore, (A.24) can be straightforwardly obtained by
subtracting (A.20c) (multiplied by $C^{-1}$ from the right) from (A.20a)
(multiplied by C$^{-1}$ from the left). Thus, contrary to the claims made in
[95], equations (A.12) and (A.20) are, in fact, equivalent. Nevertheless, as
claimed in [95], the system of equations (A.20) is easier to connect with
the Bethe ansatz formalism.

Indeed, using already known fact that $D_{n}$\textbf{\ =}$C^{n-1}DC^{-n}$
Eq.(A.22) can be brought into the form: 
\begin{equation}
\dot{D}_{n}=\frac{1}{2}[p_{R}D_{n+1}+p_{L}D_{n-1}-(p_{R}+p_{L})D_{n}]. 
\tag{A.25}
\end{equation}%
This result is formally in agreement with previously obtained (A.14) for the
fully symmetric case. The authors of [95] have chosen such a normalization
for probabilities $p_{R}$ and $p_{L}$ that for symmetric case $p_{R}=p_{L}=1$
(instead of $p_{R}=p_{L}=1/2).$ To restore the normally accepted condition $%
p_{R}=p_{L}=1/2$ requires only to rescale time appropriately. \ This
observation is consistent with the fact that the analog of equation (A.15)
(which plays the central role in Bethe ansatz-type calculations) obtained
with help of (A.23) is given by 
\begin{equation}
p_{L}D_{n+1}D_{n}+p_{R}D_{n}D_{n-1}=(p_{R}+p_{L})D_{n}D_{n}  \tag{A.26}
\end{equation}%
which holds true wether we choose $p_{R}=p_{L}=1$ or $p_{R}=p_{L}=1/2%
\footnote{%
It should be noted though that the authors of [95] have erroneously obtained
(e.g. see their equation (23)) \ $%
p_{R}D_{n}^{2}+p_{L}D_{n+1}^{2}=(p_{R}+p_{L})D_{n}D_{n+1}$ instead of our
(A.26).}.$ These results allow us to reobtain the $S-$matrix for the XXZ
model in a way \ already described.

\ 

\textbf{A.3.} \textbf{Steady- state and q- algebra for the deformed harmonic
oscillator}

\ 

Using (79) we write%
\begin{equation}
p_{R}DE-p_{L}ED=\zeta \left( D+E\right) .  \tag{A.27}
\end{equation}%
Let now $D=A_{1}+B_{1}a$ and $E=A_{2}+B_{2}a^{+}$ where $A_{i}$ and $B_{i}$, 
$i=1,2,$ are some $c-$numbers. Substituting these expressions back to (A.27)
we obtain the following set of equations%
\begin{equation}
\zeta (A_{1}+A_{2})-\varepsilon A_{1}A_{2}=C,  \tag{A.28a}
\end{equation}%
where $C$ is some constant to be determined below, and%
\begin{equation}
\zeta B_{1}=\varepsilon B_{1}A_{2},  \tag{A.28b}
\end{equation}%
\begin{equation}
\zeta B_{2}=\varepsilon B_{2}A_{1}.  \tag{A.28c}
\end{equation}%
From here we obtain: $A_{1}=A_{2}=A=\zeta /\varepsilon ,$ with $B_{1},$ $%
B_{2}$ being yet arbitrary c-numbers and $\varepsilon =p_{R}-p_{L}$. \ We
can determine these numbers by comparing our results with those in [61].
This allows us to select $B_{1}=B_{2}=\frac{\xi }{\sqrt{1-q}},$ $\dfrac{%
\zeta ^{2}}{\varepsilon }=\frac{\xi ^{2}}{1-q}=C,$ $q=\dfrac{p_{L}}{p_{R}}$
so that we obtain%
\begin{equation}
D=\frac{1}{1-q}+\frac{1}{\sqrt{1-q}}a,  \tag{A.29a}
\end{equation}%
\begin{equation}
E=\frac{1}{1-q}+\frac{1}{\sqrt{1-q}}a^{+}  \tag{A.29b}
\end{equation}%
and, finally,%
\begin{equation}
aa^{+}-qa^{+}a=1  \tag{A.29c}
\end{equation}%
in accord with (52d).

\bigskip

\textbf{References}

\bigskip

[1] \ \ P.Dirac, \textit{Lectures on quantum field theory},

\ \ \ \ \ \ (Yeshiva University Press, NY,1996).

[2] \ \ A.Kholodenko, Heisenberg honeycombs solve Veneziano puzzle,

\ \ \ \ \ \ \ Int.Math.Research Forum, \textbf{4} (2009) 441-509,
hep-th/0608117.

[3] \ \ A.Kholodenko, Quantum signatures of Solar System dynamic\textit{s},

\ \ \ \ \ \ \ Found.Phys. (submitted), arXiv.0707.3992.

[4] \ \ A.Kholodenko, New strings for old Veneziano amplitudes I\textit{.}

\ \ \ \ \ \ Analytical treatment, J.Geom.Phys.\textbf{55} (2005) 50-74.

[5] \ \ A.Kholodenko, New strings for old Veneziano amplitudes II.

\ \ \ \ \ \ Group-theoretic treatment\textit{, }J.Geom.Phys.\textbf{56}
(2006)1387-1432.

[6] \ \ A.Kholodenko, New strings for old Veneziano amplitudes III\textit{.}

\ \ \ \ \ \ Symplectic treatment\textit{,} J.Geom.Phys.\textbf{56} (2006)
1433-1472.

[7] \ \ A.Kholodenko, New models for Veneziano amplitudes:

\ \ \ \ \ \ Combinatorial, symplectic and supersymmetric aspects,

\ \ \ \ \ \ Int.J.Geom.Methods in Mod. Phys.\textbf{2} (2005) 563-584.

[8] \ \ P.Collins, \textit{Introduction to Regge Theory and High Energy
Physics},

\ \ \ \ \ \ (Cambridge University Press, Cambridge, UK,1977)

[9]. \ D. Bardin and G. Passarino, \textit{The Standard Model in the Making},

\ \ \ \ \ \ (Clarendon Press, Oxford, UK,1999).

[10] \ N. Dorey, A spin chain from string theory, arXiv: 0805.4387.

[11] \ S. Gubser, I. Klebanov, A.Polyakov, Gauge theory correlators

\ \ \ \ \ \ \ from non-critical string theory, Phys.Lett.\textbf{\ 428B}
(1998) 105-114.

[12] \ J. Minahan, K. Zarembo, The Bethe ansatz for $\mathcal{N}$=4 super

\ \ \ \ \ \ \ Yang-Mills, JHEP 0303 (2003) 013.

[13] \ M. Kruczenski, Spin chains and string theory,

\ \ \ \ \ \ \ PRL \textbf{93} (2004) 161602, 4 pp.

[14] \ A.Cootrone, L.Martucci, J.Pons, P.Talavera, Heavy hadron

\ \ \ \ \ \ \ spectra from spin chains and strings, JHEP \textbf{05} (2007)
027.

[15] \ G. 't Hooft, Topology of the gauge condition and new confinement

\ \ \ \ \ \ \ phases in non-Abelian gauge theories,

\ \ \ \ \ \ \ Nucl.Phys \textbf{190B} (1981) 455-478.

[16] \ Y. Nambu, Strings, monopoles and gauge fields,

\ \ \ \ \ \ \ Phys.Rev.\textbf{D10 }(1974) 4262-4268.

[17] \ \ T. Suzuki, I. Yotsuyanagi, \ Possible evidence for Abelian dominance

\ \ \ \ \ \ \ \ in quark confinement, Phys.Rev. \textbf{D 42} (1990)
4257-4260.

[18] \ \ J. Stack, S. Neiman, R.Wensley, String Tension from Monopoles

\ \ \ \ \ \ \ \ in SU(2) Lattice Gauge Theory, Phys Rev.\textbf{D 50} (1994)
3399-3405.

[19] \ \ Y. Cho, Restricted gauge theory, Phys. Rev.\textbf{D 21} (1980)
1080-1088.

[20] \ \ Y. Cho, D. Pak, Monopole condensation in SU(2) QCD,

\ \ \ \ \ \ \ \ Phys.Rev.\textbf{D 65} (2002) 074027.

[21] \ \ \ \ K-I. Kondo, Gauge-invariant gluon mass, infrared Abelian
dominance,

\ \ \ \ \ \ \ \ and stability of magnetic vacuum, Phys. Rev. \textbf{D 74}
(2006) 125003.

[22] \ \ \ K-I. Kondo, A. Ono, A. Shibata, T. Shinohara, T. Murakami,

\ \ \ \ \ \ \ \ Glueball mass from quantized knot solitons and

\ \ \ \ \ \ \ \ gauge-invariant gluon mass, J.Phys. \textbf{A 39} (2006)
13767--13782.

[23]\ \ \ K-I. Kondo, Magnetic monopoles and center vortices

\ \ \ \ \ \ \ \ as gauge-invariant topological defects simultaneously
responsible

\ \ \ \ \ \ \ \ for confinement, arXiv: 0802.3829.

[24] \ \ L. Faddeev, Knots as possible excitations of the quantum

\ \ \ \ \ \ \ \ Yang-Mills fields, arXiv: 0805.1624.

[25] \ \ D. Auckley, L. Kapitanski, J. Speight, Geometry and analysis

\ \ \ \ \ \ \ \ in nonlinear sigma models, St.Petersburg Math. J. \textbf{18}
(2007) 1-19.

[26] \ \ \ Y.Cho, D.Pak, P.Zhang, New interpretation of Skyrme theory,

\ \ \ \ \ \ \ \ \ arXiv: hep-th/0404181

[27] \ \ \ A.Kholodenko, Veneziano amplitudes, spin chains and

\ \ \ \ \ \ \ \ \ Abelian reduction of QCD, arXiv: 0810.0250

[28] \ \ \ A.Polychronakos, Exact spectrum of SU(n) spin chain with

\ \ \ \ \ \ \ \ \ inverse square exchange\textit{, }Nucl.Phys. \textbf{B419}
(1994) 553-566.

[29] \ \ \ \ R.Stanley, \textit{Combinatorics and commutative algebra}

\ \ \ \ \ \ \ \ \ (Birkh\"{a}user, Boston 1996).

[30] \ \ \ S.Ghorpade and G.Lachaud, Hyperplane sections of Grassmannians

\ \ \ \ \ \ \ \ \ and the number of MDS linear codes, Finite Fields \&

\ \ \ \ \ \ \ \ Their Applications \textbf{7 (}2001) 468-476\textbf{.}

[31] \ \ \ R.Stanley, Enumerative combinatorics, Vol.1

\ \ \ \ \ \ \ \ \ (Cambridge University Press, Cambridge, U.K.1999).

[32] \ \ \ S.Mohanty, \textit{Lattice path counting and applications},

\ \ \ \ \ \ \ \ \ (Academic Press, NY, 1979).

[33] \ \ \ R.Bott and L.Tu, \textit{Differential forms in algebraic topology}%
,

\ \ \ \ \ \ \ \ \ (Springer-Verlag, Berlin, 1982).

[34] \ \ \ J.Schwartz, \textit{Differential geometry and topology},

\ \ \ \ \ \ \ \ \ (Gordon and Breach, Inc., NY, 1968).

[35] \ \ \ \ M.Stone, Supersymmetry and quantum mechanics of spin,

\ \ \ \ \ \ \ \ \ \ Nucl.Phys. \textbf{B 314} (1989) 557-569.

[36] \ \ \ \ O.Alvarez, I Singer and P.Windey, Quantum mechanics and

\ \ \ \ \ \ \ \ \ \ the\ geometry of the Weyl character formula\textit{,}

\ \ \ \ \ \ \ \ \ \ Nucl.Phys.\textbf{B 337} (1990) 467-482.

[37] \ \ \ \ A.Polyakov, \textit{Gauge fields and strings},

\ \ \ \ \ \ \ \ \ \ (Harwood Academic Publ., NY, 1987).

[38] \ \ \ \ \ H. Frahm, \textit{Spectrum of a spin chain with invese square
exchange},

\ \ \ \ \ \ \ \ \ \ J.Phys.\textbf{A 26} (1993) L473-479.

[39] \ \ \ \ A.Polychronakos, Generalized statistics in one dimension,

\ \ \ \ \ \ \ \ \ \ hep-th/9902157.

[40] \ \ \ \ A.Polychronakos, Physics and mathematics of Calogero particles,

\ \ \ \ \ \ \ \ \ \ hep-th/0607033.

[41] \ \ \ \ K.Hikami, Yangian symmetry and Virasoro character in a lattice

\ \ \ \ \ \ \textit{\ }\ \ \ spin \ system with long -range interactions,

\ \ \ \ \ \ \ \ \ \ Nucl.Phys.\textbf{B 441} (1995) 530-568.

[42] \ \ \ \ E.Melzer, The many faces of a character\textit{,}
hep-th/9312043.

[43] \ \ \ \ R.Kedem, B.McCoy and E.Melzer, The sums of Rogers, Schur and

\ \ \ \ \ \ \ \ \ \ Ramanujian and the Bose-Fermi correspondence in 1+1

\ \ \ \ \ \ \ \ \ \ dimensional quantum field theory, hep-th/9304056.

[44] \ \ \ \ A.Tsvelik, \textit{Quantum Field Theory in Condensed Matter
Physics},

\ \ \ \ \ \ \ \ \ \ (Cambridge University Press, Cambridge, U.K., 2003).

[45] \ \ \ \ \ J.Goldman and J.C.Rota, The number of subspaces of a vector
space\textit{,}

\ \ \ \ \ \ \ \ \ \ in Recent Progress in Combinatorics,

\ \ \ \ \ \ \ \ \ \ pp.75-83. (Academic Press, NY,1969).

[46] \ \ \ \ V.Kac and P.Cheung, \textit{Quantum calculus}

\ \ \ \ \ \ \ \ \ \ (Springer-Verlag, Berlin , 2002).

[47] \ \ \ \ \ G.Andrews, \textit{The theory of partitions}

\ \ \ \ \ \ \ \ \ \ \ (Addison-Wesley Publ.Co., London,1976).

[48] \ \ \ \ \ D.Galetti, Realization of the q-deformed harmonic oscillator:

\ \ \ \ \ \ \ \ \ \ \ Rogers-Szego and Stiltjes-Wiegert polynomials,

\ \ \ \ \ \ \ \ \ \ \ Brazilian Journal of Physics \textbf{33} (2003)148-157.

[49] \ \ \ \ \ M.Chaichian, H.Grosse and P.Presnajer, Unitary representations

\ \ \ \ \ \ \ \textit{\ \ \ \ }of the q-oscillator algebra, J.Phys.A\textbf{%
\ 27} (1994) 2045-2051.

[50] \ \ \ \ \ R. Floreani and L.Vinet, Q-orthogonal polynomials and the

\ \ \ \ \ \ \ \ \ \ \ oscillator quantum group, Lett.Math.Phys. \textbf{22}
(1991) 45-54.

[51] \ \ \ \ \ H.Karabulut, Distributed \ Gaussian polynomials as
q-oscillator

\ \ \ \ \ \ \ \ \ \ \ eigenfunctions, \ J.Math.Phys.\textbf{47} (2006)
013508.

[52] \ \ \ \ \ \ A.Macfarlane, On q-analogues of the quantum harmonic
oscillator

\ \ \ \ \ \ \ \ \ \ \ \ and \ the quantum group\textit{\ SU(2)}$_{q},$
J.Phys.A\textbf{\ 22} (1989) 4581-4588.

[53] \ \ \ \ \ \ H.Karabulut and E.Siebert, Distributed \ Gaussian
polynomials

\ \ \ \ \ \ \ \ \ \ \ \ and associated \ Gaussian quadratures\textit{,}

\ \ \ \ \ \ \ \ \ \ \ \ J.Math.Phys.\textbf{38} (1997) 4815-4831

[54] \ \ \ \ \ \ A.Atakishiev and Sh.Nagiyev, On the Rogers-Szego
polynomials,

\ \ \ \ \ \ \ \ \ \ \ \ J.Phys.\textbf{A 27} (1994) L611-615.

[55] \ \ \ \ \ \ G.Gasper and M.Rahman, \textit{Basic hypergeometric series},

\ \ \ \ \ \ \ \ \ \ \ \ (Cambridge University Press, Cambridge, U.K, 1990).

[56] \ \ \ \ \ \ \ \ \ R.Koekoek and R.Swarttouw, The Askey- scheme of
hypergeometric

\ \ \ \ \ \ \ \ \ \ \ \ orthogonal polynomials and its q-analogs\textit{, }%
arXiv:math/9602214.

[57] \ \ \ \ \ \ M.Ismail, \textit{Classical and Quantum Orthogonal
Polynomials of One}

\ \ \ \ \ \ \ \ \ \ \ \textit{Variable} (Cambridge University Press,
Cambridge, U.K, 2005).

[58] \ \ \ \ \ \ T.Nagao and T.Sasamoto,\textit{\ }Asymmetric simple
exclusion process

\ \ \ \ \ \ \ \ \ \ \ \ and modified random matrix ensembles,

\ \ \ \ \ \ \ \ \ \ \ \ Nucl.Phys.\textbf{B 699} (2004) 487-502.

[59] \ \ \ \ \ \ R.Stinchcombe and G.Schutz, Application of operator algebras

\ \ \ \ \ \ \ \ \ \ \ \ to\textit{\ \ }stochastic dynamics and Heisenberg
chain,

\ \ \ \ \ \ \ \ \ \ \ \ PRL \textbf{75} (1995) 140-144.

[60] \ \ \ \ \ \ T.Sasamoto, One-dimensional partially asymmetric simple

\ \ \ \ \ \ \ \ \ \ \ \ exclusion process with open boundaries: orthogonal

\ \ \ \ \ \ \ \ \ \ \ \ polynomials approach, J.Phys.\textbf{A 32} (1999)
7109-7131.

[61] \ \ \ \ \ \ R.Blythe, M.Ewans, F.Colaiori and F.Essler, Exact solution

\ \ \ \ \ \ \ \ \ \ \ \ of a partially \ asymmetric exclusion model using a
deformed

\ \ \ \ \ \ \ \ \ \ \ \ oscillator algebra\textit{, }J.Phys.\textbf{A 33}
(2000) 2313-2332.

[62] \ \ \ \ \ \ B.Derrida, M.Ewans, V.Hakim and V.Pasquer, Exact solution

\ \ \ \ \ \ \ \ \ \ \ \ of a 1d asymmetric exclusion model using a matrix
formulation,

\ \ \ \ \ \ \ \ \ \ \ \ J.Phys \textbf{A 26} (1993) 1493-1517.

[63] \ \ \ \ \ \ B.Derrida and K.Malick, Exact diffusion constant for the
one-

\ \ \ \ \ \ \ \ \ \ \ \ dimensional \ partially asymmetric exclusion model,

\ \ \ \ \ \ \ \ \ \ \ \ J.Phys.\textbf{A 30} (1997) 1031-1046.

[64] \ \ \ \ \ \ M.Kardar, G.Parisi and Yi-Ch.Zhang, Dynamic scaling of
growing

\ \ \ \ \ \ \ \ \ \ \ \ interfaces, PRL \textbf{56} (1986) 889-892.

[65] \ \ \ \ \ \ \ T.Sasamoto, S.Mori and M.Wadati, One-dimensional
asymmetric

\ \ \ \ \ \ \ \ \ \ \ \ \ exclusion model with open boundaries,

\ \ \ \ \ \ \ \ \ \ \ \ \ J.Phys.Soc.Japan \textbf{65} (1996) 2000-2008.

[66] \ \ \ \ \ \ \ T. Oliviera, K.Dechoum, J.Redinz and F. Aarao Reis,
Universal

\ \ \ \ \ \ \ \ \ \ \ \ \ and nonuniversal \ features of the crossover from
linear to

\ \ \ \ \ \ \ \ \ \ \ \ \ nonlinear interface growth,

\ \ \ \ \ \ \ \ \ \ \ \ \ Phys.Rev. \textbf{E 74} (2006) 011604.

[67] \ \ \ \ \ \ \ \ A.Lazarides, Coarse-graining a restricted solid-on
-solid model,

\ \ \ \ \ \ \ \ \ \ \ \ \ \ Phys.Rev.\textbf{E 73} (2006) 041605.

[68] \ \ \ \ \ \ \ \ D.Huse, Exact exponents for infinitely many new
multicritical

\ \ \ \ \ \ \ \ \ \ \ \ \ \ points, \ Phys.Rev.\textbf{B 30} (1984)
3908-3917.

[69] \ \ \ \ \ \ \ \ D.Friedan, Z.Qui and S.Shenker, Conformal
invariance,unitarity

\ \ \ \ \ \ \ \ \ \ \ \ \ \ and critical exponents in two dimensions,

\ \ \ \ \ \ \ \ \ \ \ \ \ \ PRL \textbf{52} (1984) 1575-1580.

[70] \ \ \ \ \ \ \ \ J.de\ Gier and F.Essler, Exact spectral gaps of the
asymmetric

\ \ \ \ \ \ \ \ \ \ \ \ \ \ exclusion process with open boundaries\textit{,}

\ \ \ \ \ \ \ \ \ \ \ \ \ J.Sat.Mech.(2006) P12011\textit{.}

[71] \ \ \ \ \ \ \ \ V.Pasquer and H.Saleur, Common structures between finite

\ \ \ \ \ \ \ \ \ \ \ \ \ \ systems \ and conformal field theories through
quantum groups,

\ \ \ \ \ \ \ \ \ \ \ \ \ \ Nucl.Phys. \textbf{B 330} (1990) 523-547.

[72] \ \ \ \ \ \ \ \ \ C.Gomez, m.Ruiz-Altaba and G.Sierra,

\ \ \ \ \ \ \ \ \ \ \ \ \ \ \textit{Quantum Groups in Two-Dimensional Physics%
},

\ \ \ \ \ \ \ \ \ \ \ \ \ \ (Cambridge University Press, Cambridge, UK,1996).

[73] \ \ \ \ \ \ \ \ \ L.Faddeev and O.Tirkkonen, Connections of the
Liouville

\ \ \ \ \ \ \ \ \ \ \ \ \ \ model and XXZ spin chain, Nucl.Phys.\textbf{B 453%
} (1995) 647-669.

[74] \ \ \ \ \ \ \ \ \ A.Kholodenko, \ Kontsevich-Witten model from \ 2+1
gravity:

\ \ \ \ \ \ \ \ \ \ \ \ \ \ \ \ New exact combinatorial solution,
J.Geom.Phys.\textbf{43} (2002) 45-91

[75] \ \ \ \ \ \ \ \ \ P.Forrester, Vicious random walkers in the limit of a
large

\ \ \ \ \ \ \ \ \ \ \ \ \ \ \ number of walkers\textit{, }J.Stat.Phys. 
\textbf{56} (1989) 767-784.

[76] \ \ \ \ \ \ \ \ \ T.Sasamoto, Fluctuations of the one-dimensional
asymmetric

\ \ \ \ \ \ \ \ \ \ \ \ \ \ \ exclusion process using random matrix
techniques,

\ \ \ \ \ \ \ \ \ \ \ \ \ \ \ J.Stat.Mech. (2007) P07007.

[77] \ \ \ \ \ \ \ \ \ \ A.Mukherjee and S.Mukhi, c=1 matrix models:
equivalences

\ \ \ \ \ \ \ \ \ \ \ \ \ \ \ and open-closed string duality, JHEP \textbf{%
0510} (2005) 099.

[78] \ \ \ \ \ \ \ \ \ J.Distler and C.Vafa, A critical matrix model at c=1,

\ \ \ \ \ \ \ \ \ \ \ \ \ \ \ Mod.Phys.Lett.\textbf{A 6} (1991) 259-267.

[79] \ \ \ \ \ \ \ \ \ \ D.Ghoshal and C.Vafa, c=1 strting as the
topological theory

\ \ \ \ \ \ \ \textit{\ \ \ \ \ \ \ \ }on a conifold\textit{, }Nucl.Phys. 
\textbf{B 453} (1995) 121-128.

[80] \ \ \ \ \ \ \ \ \ \ D.Huse and M.Fisher, Commensurate melting, domain
walls\textit{,}

\ \ \ \ \ \ \ \textit{\ \ \ \ \ \ \ \ }and dislocations, Phys.Rev.\textbf{B29%
} (1984) 239-248.

[81] \ \ \ \ \ \ \ \ \ M.Gaudin, \textit{La Function D'onde de Bethe,}

\ \ \ \ \ \ \ \ \ \ \ \ \ \ \ (Masson, Paris, 1983).

[82] \ \ \ \ \ \ \ \ \ \ D.Grabiner, Brownian motion in a Weyl chamber,

\ \ \ \ \ \ \ \ \ \ \ \ \ \ \ \ non -colliding particles, and random
matrices,

\ \ \ \ \ \ \ \ \ \ \ \ \ \ \ \ arXiv: math.RT/9708207.

[83] \ \ \ \ \ \ \ \ \ \ C. Krattenhaler, Asymptotics for random walks in
alcoves\textit{\ }

\ \ \ \ \ \ \ \ \ \ \ \ \ \ \ \ of affine Weyl groups, arXiv: math/0301203.

[84] \ \ \ \ \ \ \ \ \ \ \ S.de Haro, Chern-Simons theory, 2d Yang -Mills,%
\textit{\ }

\ \ \ \ \ \ \ \ \ \ \ \ \ \ \ \ and Lie algebra wanderers, Nucl.Phys. 
\textbf{B730} (2005) 312-351.

[85] \ \ \ \ \ \ \ \ \ \ \ M.Mehta, \textit{Random Matrices}

\ \ \ \ \ \ \ \ \ \ \ \ \ \ \ \ (Elsevier, Amsterdam, 2004).

[86] \ \ \ \ \ \ \ \ \ \ \ J.Maldacena, G.Moore, N.Seiberg and D.Shih,

\ \ \ \ \ \ \ \ \ \ \ \ \ \ \ \ Exact vs. semiclassical target space of the
minimal string\textit{,}

\ \ \ \ \ \ \ \ \ \ \ \ \ \ \ \ JHEP \textbf{0410} (2004) 020.

[87] \ \ \ \ \ \ \ \ \ \ \ K.Okuyama, D-brane amplitudes in topological
string on

\ \ \ \ \ \ \ \ \ \ \ \ \ \ \ \ \ conifold\textit{, }Phys.Lett.\textbf{B 645}
(2007) 275-280.

[88] \ \ \ \ \ \ \ \ \ \ \ \ M.Tierz, Soft matrix models and Chern-Simons
partition

\ \ \ \ \ \ \ \ \ \ \ \ \ \ \ \ \ functions, Mod.Phys. \textbf{A 19} (2004)
1365-1378.

[89] \ \ \ \ \ \ \ \ \ \ \ \ A.Varchenko, \textit{Special Functions, KZ Type
Equations,}

\ \ \ \ \ \ \ \ \ \ \ \ \ \ \ \ \textit{\ and\ Representation Theory}

\ \ \ \ \ \ \ \ \ \ \ \ \ \ \ \ \ (AMS Publishers, Providence, RI, 2003).

[90] \ \ \ \ \ \ \ \ \ \ \ \ P.Etingof, I.Frenkel and A.Kirillov Jr., 
\textit{Lectures on}

\ \ \ \ \ \ \ \ \ \ \ \ \ \ \ \ \ \textit{Representation \ Theory and
Knizhnik-Zamolodchikov}

\ \ \ \ \ \ \ \ \ \ \ \ \ \ \ \ \ \textit{Equations},\ (AMS Publishers,
Providence, RI,1998).

[91] \ \ \ \ \ \ \ \ \ \ \ \ R. Blythe and M.Evans, Nonequilibrium steady
states of\textit{\ }

\ \ \ \ \ \ \ \ \ \ \ \ \ \ \ \ \ \ \ matrix-product form: a solver guide,

\ \ \ \ \ \ \ \ \ \ \ \ \ \ \ \ \ \ J.Phys.A \textbf{40} (2007) R333.

[92] \ \ \ \ \ \ \ \ \ \ \ \ G.Sch\"{u}tz, Exact solution of the master
equation for the\textit{\ }

\ \ \ \ \ \ \ \ \ \ \ \ \ \ \ \ \ \ \ \ \ asymmetric exclusion process,
J.Stat.Phys. \textbf{88} (1997) 427-445.

[93] \ \ \ \ \ \ \ \ \ \ \ \ \ \ L.Gwa and H.Spohn, Bethe solution for the
dynamical-scaling

\ \ \ \ \ \ \ \ \ \ \ \ \ \ \ \ \ \ exponent of the noisy Burgers equation%
\textit{,}

\ \ \ \ \ \ \ \ \ \ \ \ \ \ \ \ \ \ Phys.Rev.A \textbf{46} (1992) 844-854.

[94]\ \ \ \ \ \ \ \ \ \ \ \ \ \ F.Alcaraz, M.Barber, M.Batchelor, R.Baxter
and G.Quispei,

\ \ \ \ \ \ \ \ \textit{\ \ \ \ \ \ \ \ \ \ }Surface exponents of the
quantum XXZ, Ashkin-Teller and

\ \ \ \ \ \ \ \ \ \ \ \ \ \ \ \ \ \ Potts models, J.Phys.A \textbf{20}
(1987) 6397-6416.

[95] \ \ \ \ \ \ \ \ \ \ \ \ T.Sasamoto and M.Wadati, Dynamic matrix product
ansatz

\ \ \ \ \ \ \ \ \ \ \ \ \ \ \ \ \ \ and Bethe ansatz equation for asymmetric
exclusion process

\ \ \ \ \ \ \ \ \ \ \ \ \ \ \ \ \ \ with periodic boundary, J.Phys.Soc.Jpn. 
\textbf{66} (1997) 279-282.

\textit{\ \ \ \ \ \ }

\bigskip

\ \ 

\end{document}